%% file: mainTSE_new.tex
\documentclass[10pt,journal,cspaper,compsoc]{IEEEtran}
\usepackage{color}
\usepackage{pifont}

\usepackage{subcaption}
\usepackage{graphicx}
\usepackage{listings}
\usepackage{booktabs}
\usepackage{array}
\usepackage{amsmath}
\usepackage{balance}
\usepackage{multirow}
\usepackage{paralist}
\usepackage{hyperref}
\usepackage{wrapfig}
\newcolumntype{L}[1]{>{\raggedright\let\newline\\\arraybackslash}p{#1}}
\definecolor{bg}{rgb}{0.25,0,0.1}

\usepackage[hang]{footmisc}
\usepackage{lipsum}

\begin{document}

\title{Leveraging Historical Associations between Requirements and Source Code to Identify Impacted Classes}

\author{\IEEEauthorblockN{Davide Falessi\IEEEauthorrefmark{1}~ \IEEEmembership{Senior Member,~IEEE}, Justin Roll\IEEEauthorrefmark{1}~, Jin L.C. Guo\IEEEauthorrefmark{2}~, and Jane Cleland-Huang\IEEEauthorrefmark{3}~\IEEEmembership{Member,~IEEE} }\\
\IEEEauthorblockA{\IEEEauthorrefmark{1}California Polytechnic State University, San Luis Obispo, CA, 93407\\
    dfalessi@calpoly.edu; jroll@calpoly.edu  }\\ 
\IEEEauthorblockA{\IEEEauthorrefmark{2}School of Computer Science\\
	McGill University, Montreal, Canada\\
	jguo@cs.mcgill.ca}  \\
\IEEEauthorblockA{\IEEEauthorrefmark{3}Computer Science and Engineering\\
	University of Notre Dame, South Bend, IN\\
	JaneClelandHuang@nd.edu}  
}

% The paper headers
\markboth{IEEE TRANSACTIONS ON SOFTWARE ENGINEERING, VOL. XX, NO. X, XXXXXXX 2017}%
{Shell \MakeLowercase{\textit{et al.}}: Bare Demo of IEEEtran.cls for Computer Society Journals}
\IEEEcompsoctitleabstractindextext{%

\begin{abstract}
As new requirements are introduced and implemented in a software system, developers must identify the set of source code classes which need to be changed. Therefore, past effort has focused on predicting the set of classes impacted by a requirement. In this paper, we introduce and evaluate a new type of information based on the intuition that the set of requirements which are associated with historical changes to a specific class are likely to exhibit semantic similarity to new requirements which impact that class. This new Requirements to Requirements Set (R2RS) family of metrics captures the semantic similarity between a new requirement and the set of existing requirements previously associated with a class. The aim of this paper is to present and evaluate the usefulness of R2RS metrics in predicting the set of classes impacted by a requirement. We consider 18 different R2RS metrics by combining six natural language processing techniques to measure the semantic similarity among texts (e.g., VSM) and three distribution scores to compute overall similarity (e.g., average among similarity scores).  We evaluate if R2RS is useful for predicting impacted classes in combination and against four other families of metrics that are based upon temporal locality of changes, direct similarity to code, complexity metrics, and code smells. Our evaluation features five classifiers and 78 releases belonging to  four large open-source projects, which result in over 700,000 candidate impacted classes. 
Experimental results show that leveraging R2RS information increases the accuracy of predicting impacted classes practically by an average of more than 60\% across the various classifiers and projects.
\end{abstract}

}
\maketitle

%\abstract{Abstract}
\input{Introduction}
\input{Approach}

\input{ExperimentalDesign}

\input{Results}

\input{Threats}
\input{Related}
\input{Conclusion}

\section*{Acknowledgments}
Davide Falessi was partially supported by the Cal Poly SURP grant "Enhancing Software Impact Analysis via Semantic Requirements Analysis'' and the Cal Poly RSCA grant  "Software Change Predictability via Semantic Requirements Analysis''.
The Notre Dame researchers were partially funded in this project by the US National Science Foundation Grant CCF-1319680.

\bibliographystyle{IEEEtran}
\bibliography{bibliography}
\input{Appendix}

\input{Biography}
\end{document}

%% file: Introduction.tex
\section{Introduction}
Software projects are commonly developed incrementally as new requirements are introduced over time across multiple releases \cite{DBLP:conf/icse/Murphy16, DBLP:conf/icse/StahlB16}. Requirements are specified, formally or informally, using a range of natural language formats. For example, requirements may be written in a traditional way describing what the system `shall' or `must' do, as user stories, or as informal and unstructured feature requests logged and managed in an issue tracking system \cite{dybaa2008empirical}.  In order to implement a requirement, developers must first identify the set of source code classes, referred to as the \emph{impact set}, that would be affected by the requirement's introduction.  This task, referred to as \emph{change impact analysis} \cite{DBLP:conf/icsm/Bohner96}, is relatively simple when the project is small or when developers are knowledgeable about the code base but becomes more challenging as the system grows in size or when developers leave the project.   A newly hired programmer must often spend a great deal of time pouring over design documentation, F.A.Q.s, and project wikis in order to understand how a new feature impacts the system \cite{DBLP:conf/se/MaderE16, Wetzlmaier2015}.

The goal of impact analysis is to help developers identify an accurate impact set, thereby aiding the task of effort estimation, enabling better requirements prioritization and release decisions, and reducing the need for post-release modifications. The impact set may include sets of  classes, methods, or  sections of code, which should be modified to accommodate the change. In this paper we focus on Java files and we do not consider changes related to refactoring. The initial impact set is often extended by the developer during the implementation process as additional, and potentially unintended consequences of the change are discovered. 

The task of identifying an initial impact set can be performed manually using a programmer's existing knowledge of the system or through leveraging supporting tools to help programmers search for impacted parts of the code \cite{Wetzlmaier2015}. Automated techniques for detecting an initial set of impacted classes include information retrieval methods such as the Vector Space Model (VSM) or Latent Semantic Analysis (LSA)  \cite{DBLP:journals/tse/AntoniolCCLM02,DBLP:journals/tosem/PoshyvanykGM12, cleland2014software, DBLP:journals/tse/HayesDS06} to search for code classes which are textually and/or semantically similar to the new requirement.  Other techniques focus on identifying classes which are more likely to change based on their complexity \cite{DBLP:journals/smr/OlagueEMD08} or their history of change \cite{moonen:2016:exploring, ZimmermannTSE2005}.

Once an initial impact set is discovered, other techniques can be used to extend the set. Software objects can be modeled as a dependency matrix or graph, from which both direct and indirect impacts of a change can be identified \cite{bohner2002extending, Weiser:1981,Gallagher83912}. Solutions are based upon call graphs representing function calls extracted from code, execution traces capturing runtime execution of the code, Program Dependency Graphs representing dependencies between certain code fragments, Probabilistic Models using Markov Chains or Bayesian systems to check how likely a change is to modify other code fragments and mining and leveraging associations between co-evolving artifacts \cite{moonen:2016:exploring, lehnert2011taxonomy}. The different types of information produced by each of these techniques can be used in prediction models to identify the impact set.

\subsection{Aim, method, and research questions}
The likelihood of a class being impacted by a requirement may depend on several different factors such as the size or the frequency of changes of the class. In this paper we build on the intuition that a software system is likely structured, intentionally or spontaneously to support separation of concerns \cite{Falessi:2011:DTS:1978802.1978812}. In other words, a class is impacted by a requirement according to the semantics of the requirement. We use the natural language text as proxy of the semantic expressed in a requirement. We also use the set of requirements which are associated with historical changes to a specific class as a proxy of the semantics of the class. 

The aim of this paper is to present and evaluate the usefulness of a new family of metrics, which we entitle \emph{Requirements to Requirements Set} (R2RS), for predicting the set of classes impacted by a requirement. R2RS builds on the intuition that the set of requirements which are associated with historical changes to a specific class are likely to exhibit semantic similarity to new requirements which impact that class. We consider 18 variants of R2RS metrics by combining six natural language processing (NLP) techniques to measure the semantic similarity among texts (e.g., VSM) and three distribution scores to compute overall similarity (e.g., average across similarity scores).

Our aim is to support the work of practitioners working on industrial projects.  However, in this work we utilized Open-Source data, which is readily available. Specifically, we evaluate if the 18 R2RS metrics are useful for predicting impacted classes when used individually or in combination with 16 metrics related to temporal locality of changes, direct similarity to code, complexity metrics, and code smells. Our evaluation features five classifiers and 78 releases belonging to four large open-source projects, each project has from 1,900 to 8,119 issues, and a total of 487 new requirements, close to 4,000 java source code classes, resulting in over 700,000 candidate impacted classes. 

The best approach to understand if a family of metrics is useful for predicting a variable is to perform an exhaustive search and identify the precise combination of metrics providing the highest accuracy. Unfortunately, a fully exhaustive search is often infeasible. For instance, in our case we have 34 different metrics and an exhaustive search would have required computing the accuracy of $2^{34}=17,179,869,184$ combinations of metrics. Moreover, because the metrics may perform differently in different prediction classifiers, we would need to  test all combinations on several (i.e., five) different classifiers. Thus, if computing the accuracy of a combination of metrics on a classifier requires one second (an extremely conservative estimation), computing the accuracy of all possible metrics on five classifier would take up to 2,723 years. %(17179869184 cases * 5 classifiers * 1 sec) / (60 * 60 * 24 * 365) = 
Therefore, to evaluate the effectiveness of R2RS metrics, we apply three methods but do not attempt an exhaustive search. Each of the following research questions applies a specific method as summarized below and described in detail in subsequent sections of the paper. 

\begin{itemize}
    \item [\bf RQ1:] {\bf Do R2RS metrics provide information to predict impacted classes?}  We compute the amount of information provided by each of the 34 metrics to predict impacted classes.    
    \item [\bf RQ2:] {\bf How often are R2RS metrics selected by the classifiers to predict impacted classes?} We perform a semi-exhaustive search, by means of a standard subset evaluation in combination with a wrapper method, to identify the specific combinations of metrics that maximizes the accuracy of each classifier in each dataset. 
    \item [\bf RQ3:] {\bf What is the accuracy of single metrics families in predicting impacted classes?} We measure the accuracy of predicting impacted classes when using single families of metrics. 
    \item [\bf RQ4:] {\bf How much does using  R2RS improve the accuracy in predicting impacted classes?} We measure the accuracy of predicting impacted classes when using versus not using R2RS metrics.
\end{itemize} 

\subsection{Paper Overview}
The remainder of this paper is laid out as follows.  In Section \ref{sec:metrics} we introduce each of the five metric families, with emphasis on the novel R2RS metrics. Sections  \ref{sec:ExperimentalDesign} and \ref{sec:Results} describe the experiments we designed to address the three research questions and then discuss and analyse the results.  Section \ref{sec:Practice} discusses the implications of our results on industrial practice. Finally, from Sections \ref{sec:Threats} to \ref{sec:Conclusion} we present threats to validity, related work, and conclusions.

%% file: Approach.tex
\section{Families of Metrics}
\label{sec:metrics}

In this section we provide a detailed description of the families of metrics used in our study.

\subsection{Similarity to Class's Requirements Set (R2RS) }
\label{sec:R2RS}
\begin{figure}[!t]
   \begin{minipage}[b]{.95\linewidth}
     \centering
     \includegraphics[width=1.0\textwidth]{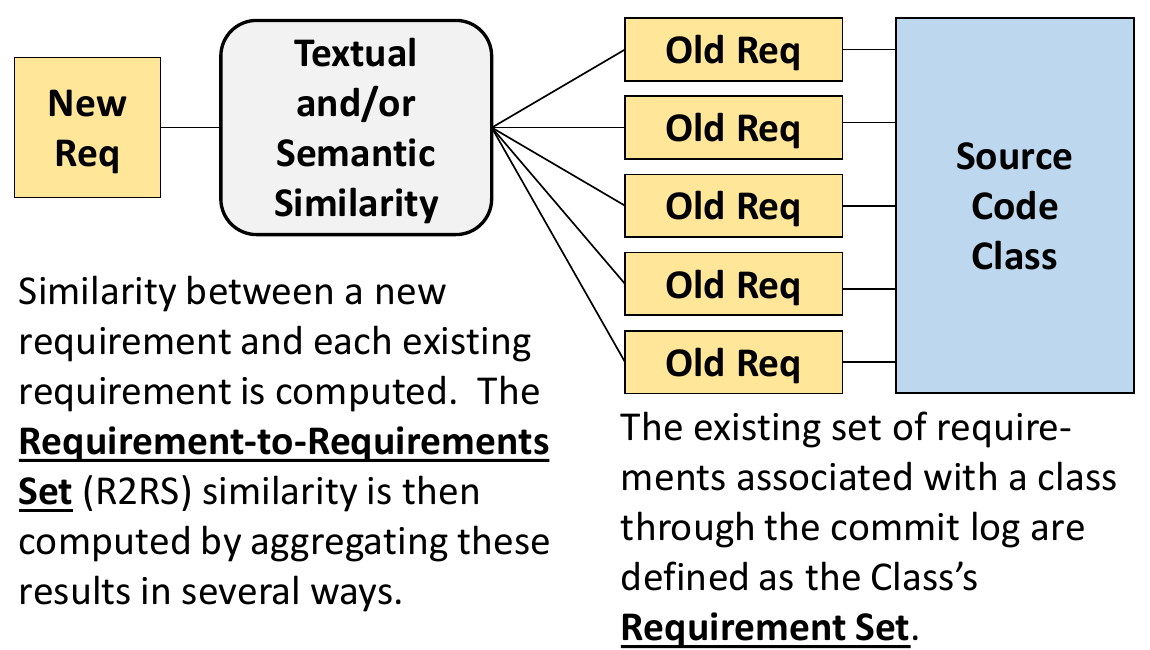}
     \hspace{-20pt}
     \subcaption{Requirements to Requirements Set ({\bf R2RS}) }\label{fig:M1}
     \hspace{30pt}
   \end{minipage}

  \begin{minipage}[b]{.95\linewidth}
     \centering
     \includegraphics[width=.8\textwidth]{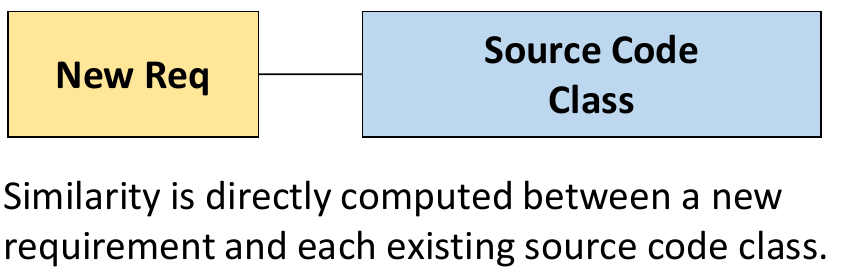}
     \hspace{-20pt}
     \subcaption{Direct Requirements to Class ({\bf R2C}) }\label{fig:M2}
     \hspace{20pt}
   \end{minipage}
    \caption{Textual similarity metrics used by the Class Change Predictor.}
   \label{fig:TextualMetrics}
\end{figure}

This novel metric family, called \textbf{R2RS}, leverages historical associations between requirements and source code to predict the impact of a new requirement. Our hypothesis is that if an existing source code class is associated with a set of previously implemented requirements through explicit references in the source code commit log, and if a new requirement is semantically similar to those prior requirements, then the new requirement is more likely to impact the same class. The approach is summarized in Figure \ref{fig:M1}. 
We illustrate the use of R2RS using an example drawn from one of our case projects, namely the \emph{Accumulo} distributed, key-value database system. Table \ref{tab:Requirements} depicts four commits associated with recently resolved new feature requests (i.e., requirements) for the Accumulo system.  Each requirement is associated with a set of classes that were created and/or modified in order to implement the requirement. Here we show only a partial listing of classes for each commit.

Individual classes tend to be modified during multiple commits.   For example, in Table \ref{tab:Requirements} we see that the class \emph{ThriftTransportKeyTest.java} was modified as part of two separate commits (\#2815 and \#3513) and is therefore associated with at least two requirements.  If these commits were the only ones under consideration, the class would have a Requirements Set of size two; however, in reality, classes tend to have several associated requirements. 

To identify the requirements set for each class we follow a series of steps. We assume that developers include the ID of the relevant requirement in the source code commit message. For example, if a requirement is created named ACCUMULO-1009, the commit associated with the requirement must include [ACCUMULO-1009] in its commit message and each commit would be evaluated using the regular expression: $.*ACCUMULO-([0-9]+).*$. This is a standard practice promoted in many open (and closed) source projects.  Furthermore, our previous study \cite{raff} showed that approximately 60\% of commits are linked to issues (e.g., requirements or bugs). Because the absence of the ticket ID in the commit message can bias results \cite{Bird:2009:FBB:1595696.1595716}, we selected projects maximizing the existence of this information. As a result we identify a set of commits associated with each requirement, a list of source code files that were associated with those commits, and the list of files present in the system at the time of the first commit of the requirement. Finally, we identify the set of requirements associated with each class. We refer to this as the class's \emph{Requirements Set}. The ultimate purpose of the requirements set is to allow us to analyze the textual similarity of a new requirement against requirements currently associated with the class.  We limit this set to the ten most recently implemented requirements associated with the class as per the dates in the commit log. We established the size of the set because we observed that the text of requirements implemented by a class lose relevance (prediction) value over time as the class evolves and implements other requirements. While the maximum set size is somewhat arbitrary, we defer the investigation of the optimal value to future studies. 

In this paper we consider 18 different R2RS metrics by combining six NLP techniques to measure the semantic similarity between requirement pairs (e.g., VSM) and three distribution scores to compute overall similarity between requirement to requirements set (e.g., average among similarity scores). The following two subsections detail the NLP techniques and distribution scores.

\begin{table*}
\addtolength{\tabcolsep}{-4pt} 
\caption{Feature Requests serve as Requirements for the Accumulo System.  Requirements are associated with Classes during the Commit process.  The total set of requirements associated with a specific class over all commits in a given time period compose that class's Requirements Set.}
\label{tab:Requirements}
\vspace{6pt}
\begin{tabular}{|L{1cm}|L{3.5cm}|L{6.6cm}|L{5.5cm}|}
\hline
{\bf Commit ID} & {\bf Feature Request (aka Requirement)} & {\bf Requirement Description \newline (Only partial text shown here)}& {\bf Modified Classes \newline (per the GitHub commit log)}\\ \hline

2181 &  Organize tables on monitor page by namespace&Improve the look and feel of the monitor by using Twitter's bootstrap. & monitor/servlets/TablesServlet.java, monitor/util/Table.java \\ \hline
2815 & Support for Kerberos client authentication. & Leverage SASL transport provided by Thrift which can speak GSSAPI, which Kerberos implements.&ThriftTransportKeyTest.java \newline ClientOpts.java,~~ more.. \\ \hline
2998 & Provide a mechanism to allow clients to wait for balance &Wait for the balancer to run and find no work to do after creating a bunch of splits.& master/Master.java \\ \hline
%1680 & Atomic add and remove of authorizations&Adding and removing an auth presents a race condition because of the two separate calls required to get, modify, then set the authorizations for a user...&shell/Shell.java, shell/commands/ DeleteAuthsCommand.java, shell/commands/DeleteAuthsCommandTest.java\\ \hline
3513&Add delegation token support for kerberos configurations. & Generate secret keys internally to Accumulo, distribute them among the nodes via ZK, and use the secret keys to create expiring passwords
that users can request and servers can validate...&ThriftTransportKeyTest.java \newline ../MapReduceClientOnDefaultTable.java  \newline ../MapReduceClientOnRequiredTable.java \newline more.. \\ \hline
\end{tabular}
\end{table*}

\subsubsection{NLP techniques}
\label{sec:Pairwise}

One of the fundamental assumptions of R2RS is that a new requirement that is textually similar to the existing requirements set for a given class is likely to impact that class.  Therefore, the first step is to compute the similarity between a new requirement $R$ and the requirements set $RS_C$ for class $C$.

The process involves three distinct steps of (1) preprocessing the text in each individual requirement, (2) computing the similarity between R and each individual requirement in $RS_C$, and (3) computing an overall score representing the similarity between $R$ and $RS_C$. \\
\noindent{\bf $\bullet$~Preprocessing:~}
As previously explained, the Requirements Set is mined from the GitHub commit log. Following standard information retrieval practices, we applied a series of pre-processing steps \cite{jivani2011comparative} which included removing non alpha-numeric characters and stemming each word to its morphological root using the Porter stemmer  \cite{jivani2011comparative}. \\
\noindent{\bf $\bullet$~Similarity between Pairs of Requirements:~} 
For experimental purposes, the similarity between two requirements was computed using a variety of NLP techniques.  These included \emph{term distribution} based methods, i.e., Vector Space Model (VSM) and Jensen Shannon Divergence (JSD), as well as methods that take sentence syntax and word semantics into consideration, i.e., Greedy Comparator,  Optimum Comparator, Corley Mihalcea Comparator, and Bleu Comparator. A brief description of each approach is provided in Table \ref{tab:ReqToClassAssociationMethods}. Each NLP technique produces a score representing the similarity between each pair of requirements. All scores were computed using \emph{TraceLab} components \cite{keenan2012tracelab,DBLP:journals/ese/DitMVPC15}. TraceLab is a plug-and-play experimental environment designed to support traceability experiments.  We used existing components to compute VSM and JSD and developed a new component which made external calls to the open source \emph{Semantic Measures} java library \cite{SemanticLibrary} in order to compute the semantic-based similarity scores. 

\begin{table}[!t]
\centering
\caption{NLP techniques to measure the textual and semantic similarity among requirements.}
\label{tab:ReqToClassAssociationMethods}
\begin{tabular}{|L{1.8cm}|L{5.8cm}|}

\hline
NLP Technique Name                & Brief Description \& Tools/Refs \\ \hline
{\bf VSM}: \newline Vector Space Model         & Compares source text T1 and target text T2 as vectors in the space constructed by index terms. Frequency of terms occurrence in the texts and in the collection of documents are normally used for weighting the terms. \cite{dasgupta2013enhancing, manning2008introduction}         \\ \hline
{\bf JSD}: \newline Jensen Shannon Divergence  & Compare T1 and T2 as two probability distributions of terms. It is based on the Kullback-Leibler divergence which measures the average inefficiency in using one distribution to code for another. \cite{dasgupta2013enhancing, lee1999measures}        \\ \hline
{\bf GC}: \newline Greedy Comparator          & Term-to-term semantic similarity measures are computed first using WordNet. Each term in T1 is then paired with every term in T2 and the maximum similarity score is greedily retained. A weighted sum is then calculated as the text similarity. \cite{DBLP:conf/acl/RusLBNS13, rus2012comparison} \\ \hline
{\bf OPC}: \newline Optimum Comparator         & Term-to-term semantic similarity measures are computed first as in Greedy Comparator. Each term in T1 is then paired with the term in T2 so that global maximum similarity score is achieved. \cite{DBLP:conf/acl/RusLBNS13, rus2012comparison}           \\ \hline
{\bf CMC}: Corley Mihalcea Comparator &  Terms in T1 are compared with terms in T2 with the same Part-of-speech (POS) tags. For nouns and verbs, the semantic similarity is used while lexical similarity is used for all other POS groups. Inverse document frequency is used for weighting terms when calculate their sum. \cite{DBLP:conf/acl/RusLBNS13, Corley:2005:MSS:1631862.1631865}\\ \hline
{\bf BC}: \newline Bleu Comparator            & Originally used for automatic evaluation of Machine Translation, compares T1 and T2 based on their n-gram (size 1 to 4) coincidences. \cite{DBLP:conf/acl/RusLBNS13, papineni2002bleu} \\ \hline
\end{tabular}
\end{table}

\subsubsection{Distribution Scores}
We considered three different approaches for computing a single similarity score given the ten similarity scores that resulted by comparing the current requirement with the last ten requirements touching the current class. These approaches, called distribution scores, were chosen because they are standard indicators of a distribution: the average, max and top five (i.e., a top percentile). 

\begin{itemize}[-]
\itemsep-.1em
\item {\bf Maximum} (MAX) represents the highest NLP technique score between the new requirement and the class's Requirements Set.
\item{\bf Average} (AVG) is the average NLP technique scores between the new requirement and the class's Requirements Set. The logic here is that looking at the entire Requirements Set associated with a class provides a more holistic representation of the class's role and responsibilities and may avoid false matches that could occur in the MAX approach, if a very similar requirement had impacted the class for an obscure reason.
\item{\bf Average of Top Five} (AvgTop5) is the average NLP technique score between the new requirement and the five requirements with the highest score among class's Requirements Set. This approach provides a balance between MAX and AVG.
\end{itemize}

Altogether, the R2RS family of metrics, therefore includes six NLP techniques for computing pairwise similarity between requirements and three similarity scores for computing the association between a requirement and a class' Requirements Set. This produces 18 different approaches, defined as metrics, for computing the R2RS. 

\subsection{Requirement-to-Class Similarity (R2C)}
\label{sec:R2C}
The second metric family, \emph{Requirements to Class} (R2C), is depicted in Figure \ref{fig:M2} and represents the direct use of information retrieval methods to trace a new requirement to source code. This approach has been described extensively in the literature \cite{DBLP:conf/se/MaderE16, ICSMEMona, DBLP:journals/tse/HayesDS06, DBLP:journals/tse/AntoniolCCLM02} and is implemented in several prototypical maintenance tools \cite{DBLP:journals/spe/LuciaFOT10, DBLP:conf/icse/PandanaboyanaSYH13,DBLP:conf/re/LinLCSABBKDZ06,keenan2012tracelab}. 
Following standard information retrieval practices, each requirement and source code class were preprocessed. Additionally, class, method, and variable names are split into constituent parts. For example, \emph{timeEvent} is split into \emph{time} and \emph{event}. As developers often use meaningful names for classes, methods, and variables, and also comment the code, this textual content can be used to estimate the similarity between a requirement and a class \cite{DBLP:journals/tse/AntoniolCCLM02, DBLP:journals/tse/MirakhorliC16}. However, much of the text in the source code is not in the form of complete sentences; thus, semantic measure algorithms are more difficult to apply. Hence, for these experiments we compute requirements to code similarity using only the Vector Space Model and Jensen Shannon Divergence.  The R2C metric family therefore includes only two metrics.

\subsection{Temporal Locality of Class Changes (TLCC)}
\label{sec:TLCC}
\begin{figure}[!t]  
    \begin{minipage}[b]{.95\linewidth}
     \centering
     \includegraphics[width=1.0\textwidth]{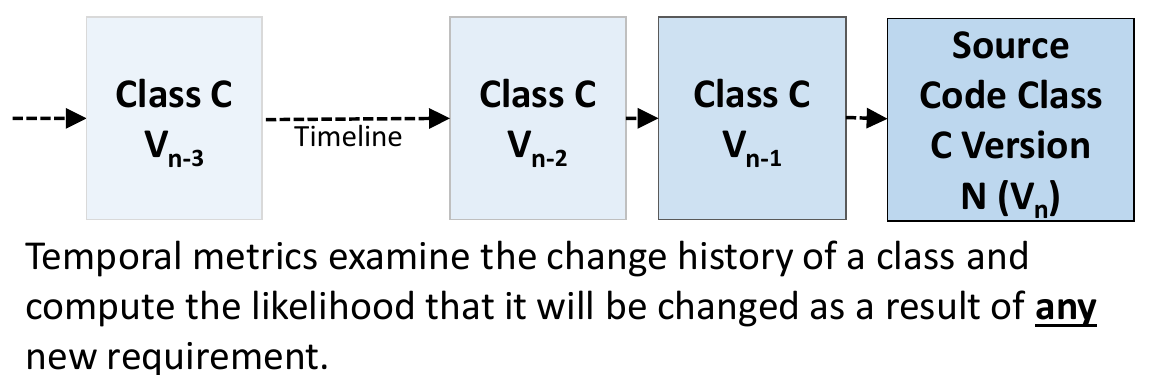}
     \hspace{-20pt}
     \hspace{30pt}
   \end{minipage}  
    \caption{Temporal locality of requirements-associated changes made to the Class ({\bf TLCC})}
   \label{fig:TempLocality}
\end{figure}

The third metric family, \emph{Temporal Locality of Class Changes} (TLCC), takes the modification history of a class into consideration. It is depicted in Figure \ref{fig:TempLocality}. Several authors have reported that change history is one of the strongest predictors of future change \cite{Graves:2000:PFI:347489.347496, DBLP:series/springer/ZimmermannNZ08} and that a class that has frequently changed in the past is more likely to change in the future and therefore more likely to be impacted by future changes \cite{DBLP:series/springer/ZimmermannNZ08}. Furthermore, research has shown that more recent commits are stronger indicators of change than less recent ones.  Kim et al. pioneered the use of \emph{temporal locality} in predicting software change, building a variety of caches to weight the importance of prior software modifications \cite{kim2007predicting}. Bernstein et al. rolled up the number of modifications for a class by month for the most recent six months and used those temporal metrics to train a defect-predicting classifier \cite{bernstein2007improving}. While TLCC works in a rather indirect way, we include it as one of our metric families because of its above-average performance in prior studies. Prior work has focused primarily on bug-fixing changes  while our emphasis in this paper is on source code changes resulting from the implementation of a new requirement.   

We compute TLCC in three different ways which represent three intuitive patterns of past source code change behavior.  In the following explanation, $N$ is the total number of requirements.  $T(i)$ is a boolean value that is 1 if the class has been touched for requirement $i$, 0 otherwise. The basic idea is that the closer $i$ is to $N$, the higher the  $T(i)$ weight is. The three metrics are defined as follows:\\

\begin{itemize}[-]
\itemsep-.1em
\item {\bf Simple Count Percent:} This divides the number of times class $C$ has been touched by the number of times the class could have been touched. 

\begin{equation}
TLCC_{SCP}(C) = \frac {\sum_{i=1}^{N} {T(i)}}{N} \
\end{equation}

\item {\bf Linear:}
This approach applies more weight to recent class modifications. If the class has been touched in the last requirement (i.e., T(i) with i=N), this will have the maximum weight, i.e., $1 / (1)$. If the class has been touched in the one before the last requirement (i.e., T(i) with i=N-1), this will have half of the maximum weight, i.e., $1 / (2)$. If the class has been touched in the two before the last requirement (i.e., T(i) with i=N-2), this will have one third of the maximum weight, i.e., $1 / (3)$.   All weights are summed and then divided by the total number of requirements ($N$).

\begin{equation}
TLCC_{Lin}(C) = \frac {\sum_{i=1}^{N} \frac{T(i)}{1 + N -i}}{N} \
\end{equation}

\item {\bf Logarithmic:} Using a linear locality may assign insufficient weight to older modifications. In this logarithmic approach the weight decrease logarithmically rather than linearly. For example, a modification at commit index 4 will be given a weight of $1 / ln(1 + 6 -4)$, the modification at commit index 5 will be given a weight of $1 / ln(1+6-5)$. Again, all weights are summed and then divided by the total number of requirements ($N$).

\begin{equation}
TLCC_{Log}(C) = \frac {\sum_{i=1}^{N} \frac{T(i)}{ln(1 + N -i)}}{N} \
\end{equation}

\end{itemize}

To better understand the calculation and interpretation of temporal locality metrics, we provide a simple example. In Table \ref{tab:TemporalLocalityTable}, six different requirements have been implemented, possibly impacting three different classes. 
For the \emph{simple} calculation, class A has been touched four different times producing a score of $\approx$ 0.67 (i.e., 4/6). Class B has been touched three times for a score of 0.5, while Class C has a score of $\approx$ 0.33. 

The linear method takes temporal locality of the class modification into consideration. Class A has been touched at its first, second, third, and fourth requirements. Class B has more recent modifications but less total modifications than Class A; however, we can see that it still has a higher score at 0.18 than class A does.  Class C has the least total modification and the most recent modifications, but it has the highest linear temporal locality score at 0.25. This exemplifies how heavily the linear temporal locality method favors recent modifications.
The logarithmic method works very similarly to the linear one, but we can see that it penalizes older activity substantially less than the linear method. Specifically, the logarithmic metric suggests class B as the most prone to future change whereas the linear metric suggests class C. This is because in the linear method class B is highly penalized by the fact that it has not been touched by the last requirement.
All three of these metrics are included in the TLCC metric family.

\subsection{Complexity via CKJM}
\begin{figure}[!t]  
    \begin{minipage}[b]{.45\linewidth}
     \centering
     \includegraphics[width=1\textwidth]{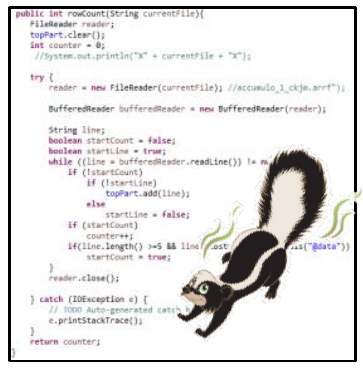}
     \hspace{-20pt}
     \subcaption{Bad Code Smells - Computed using SonarQube {\bf (SQ)}}\label{fig:M4}
     \hspace{30pt}
   \end{minipage} ~~  
    \begin{minipage}[b]{.45\linewidth}
     \centering
     \includegraphics[width=1.0\textwidth]{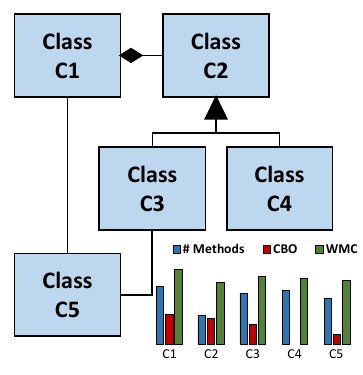}
     \hspace{-20pt}
     \subcaption{Object-Oriented Complexity metrics computed using {\bf (CKJM)} }\label{fig:M5}
     \hspace{30pt}
   \end{minipage}  
   
   \caption{Source code metrics used by the Class Change Predictor.}
   \label{fig:SourceCode}
\end{figure}

Intuitively, classes which exhibit low cohesion or expose a high number of public methods are likely to integrate multiple functions and, therefore, may need to be modified to accommodate related new requirements. Moreover, classes which are tightly coupled to many other classes may need to be changed more frequently due to ripple effects across the code. Therefore, as depicted in Figure \ref{fig:M5}, we measure several common coupling and cohesion metrics defined by Chidamber and Kemerer \cite{DBLP:journals/tse/ChidamberK94} using Spinelli's CKJM tool \cite{DBLP:journals/software/Spinellis05c}. The metrics included in the CKJM metric family are Lines of Code (LOC), Weighted Methods per Class (WMC), Depth of Inheritance Tree (DIT), Number of Children (NOC), Coupling between Object Classes (CBO), Response for a class (RFC), Lack of Cohesion of metrics (LCOM), Afferent Couplings (Ca), and Number of Public Methods (NPM).  Definitions for each metric are provided in Table \ref{tab:Codemetrics}. CKJM metrics were measured at the beginning of each project release. 

\begin{table}[]
\centering
\setlength{\tabcolsep}{3pt}
\caption{Example of Temporal Locality Result for three classes. A commit that modifies the class is indicated as X, otherwise as 0.}
\label{tab:TemporalLocalityTable}
\begin{tabular}{|c||c|c|c|c|c|c||c|c|c|}
\hline
Class & \multicolumn{6}{c||}{Requirement Touch} & \multicolumn{3}{c|}{Temporal Locality Results} \\ \cline{2-10} 
Name & 1 & 2 & 3 & 4 & 5 & 6 & Simple & Linear & Logarithmic \\ \hline
A & X & X & X & X & 0 & 0 & \textbf{0.67} & 0.16 & 0.47 \\ \hline
B & 0 & 0 & X & X & X & 0 & 0.5 & 0.18 & \textbf{0.51} \\ \hline
C & 0 & 0 & 0 & 0 & X & X & 0.33 & \textbf{0.25} & 0.24 \\ \hline
\end{tabular}
\end{table}

\begin{table}[!t]
\setlength{\tabcolsep}{1pt}
\caption{Source Code Coupling, Cohesion, and Complexity metrics}
\label{tab:Codemetrics}
\begin{tabular}{|L{2.1cm}|L{1cm}|L{5cm}|}
\hline
\centering 
	Metric&Family&Description\\ \hline

	{\bf Com}\newline Complexity  &SQ&Counts number of key-words (e.g. if, else, for, while, etc. \cite{sonarsource2013sonarqube} (Complexity) \\ \hline
	{\bf NCLOC}\newline Size in lines of code&SQ&Counts number of lines of code (excluding white space and brackets). \cite{sonarsource2013sonarqube} (Size)\\ \hline
	{\bf Viol}\newline Violations &SQ& Number of rule violations (code smells). \cite{sonarsource2013sonarqube}. (Error Propensity)\\ \hline
	{\bf WMC}: Weighted Methods per Class&CKJM&Sums number of methods per class, weighted by the cyclomatic complexity for each method \cite{cho2001component}.  (Size, Complexity)\\ \hline
	{\bf DIT}: \newline Depth of Inheritance Tree &CKJM&Measures the number of levels in a class's inheritance tree. DIT is an indicator of code-reuse \cite{cho2001component}. (Coupling)\\ \hline
	{\bf NOC}: \newline Number of Children&CKJM&Measures the number of times a class has been subclassed \cite{cho2001component}.  (Coupling)\\ \hline
	{\bf CBO}: Coupling between Object Classes&CKJM&Measures the number of non-inheritance based associations to other classes \cite{cho2001component}.  (Coupling)\\ \hline
	{\bf RFC}: \newline Response for a Class&CKJM&Counts the methods that may be invoked by a class in response to an event \cite{cho2001component}.  (Coupling)\\ \hline
	{\bf LCOM}: \newline Lack of Cohesion of metrics&CKJM&Measures the extent to which methods in a class are internally cohesive \cite{cho2001component}.  (Cohesion)\\ \hline
	{\bf Ca}: \newline Afferent Couplings&CKJM&Measures the number of direct references from other classes \cite{cho2001component}. (Coupling)\\ \hline
	{\bf NPM}: Number of Public Methods&CKJM&Measures the number of publicly exposed methods \cite{cho2001component}. (Cohesion, Coupling)\\ \hline
	\end{tabular}
\end{table}

\subsection{Bad Smells via SonarQube (SQ)}
Bad smells or code smells are the result of intentional or unintentional design and implementation choices. Examples of different smell types include Martin Fowler’s large class (aka God class) and long method \cite{1999refactoring}, low comment frequency, long if statement, and non-intuitive variable naming. Previous studies have shown that classes which are overly complex, poorly written, or exhibit `bad smells' are more likely to be refactored during a change to an associated class or generally require modifications when new functionality is introduced \cite{khomh2009exploratory, arvanitou2017method, zhou2009examining}. Therefore, as depicted in Figure \ref{fig:M4}, we measured code smells for each class using SonarQube\footnote{\url{www.sonarqube.org}}. SonarQube was selected over other static analyzers such as JDeodorant, PMD, InCode because it is free, has the ability to count hundreds of different types of smells, and was used in our prior studies \cite{DBLP:conf/icsm/FalessiV15}. SonarQube analysis was run against the commit immediately prior to each new version of a project.
%In order to count the smells in each class we had to checked out all versions of the code and run the SonarQube analysis.

%% file: ExperimentalDesign.tex
\section{Experimental Design}
\label{sec:ExperimentalDesign}
As previously explained, it was not possible to conduct an exhaustive search across all combinations of metrics in order to evaluate the effectiveness of R2RS metrics. Therefore, we proceeded to apply three complementary methods as described in the following subsections. 

\subsection{RQ1: Do R2RS metrics provide information to predict impacted classes?}
One of the most widely applied approaches for analyzing whether a metric is useful for predicting a variable is measuring the amount of information content provided by this metric \cite{Gray:1990}.
%Information gain \cite{Gray:1990} measures the amount of information in bits about the class prediction, if the only information available is the presence of a feature and the corresponding class distribution. 
Concretely, it measures the expected reduction in entropy (uncertainty associated with a random feature) \cite{Mitchell:1997}. In general, information gain ratio (IGR) is preferred over information gain because it is better at ranking metrics with a large number of distinct values \cite{harris2002information}.  

Furthermore, IGR provides advantages over prediction accuracy metrics such as precision and recall, because IGR is independent from the classifier, balancing, validation, and specific accuracy metric used. Moreover, IGR is much faster to compute than accuracy metrics. However, while relative IGR values can be easily interpreted (i.e., the higher the better), the exact meaning of individual IGR values is hard to interpret.

Note that, in our context, other widely used standard correlation metrics, such as Spearman \cite{Zimmermann:2007:PDE:1268984.1269057, Nagappan:2006:MMP:1134285.1134349}  are not suitable for two important reasons.  First, they require the variables to be in an interval scale \cite{briand1996application, stevens1946theory}, and second, they measure the degree of correlation which is orthogonal to the amount of information provided to support prediction. In contrast, IGR requires the variables to be categorical and reliably ranks metrics according to the amount of information they provide to support prediction.  For these reasons, IGR has been extensively used in the context of feature selection \cite{karegowda2010comparative}.

Information gain ratio \cite{harris2002information} is computed as the ratio between the information gain $IG$ and the intrinsic value $IV$, i.e., $IGR(Ex,a) = IG/IV$. The information gain for an attribute (i.e., metric) in the set of all attributes $Attr$ is defined as follows:
%a \url{https://en.wikipedia.org/wiki/Information_gain_ratio}}

{\footnotesize

\begin{flalign*}
\begin{split}
IG(Ex,a) =  H(Ex)- & 
\sum \limits_{v \in {values(a)}} \bigg( \frac{| \big\{ x \in{Ex|value(x,a)=v} \big\} |}{|Ex|} \cdot \\
& H\big( \big\{x\in{Ex|value(x,a)=v}\big\} \big)\bigg)
\end{split}
\end{flalign*}
}

In which $Ex$ is the set of all training examples, $value(x,a)$ with $ x\in Ex$ defines the value of a specific example $x$ for attribute $a\in Attr$, $H$ specifies the entropy, and the $values(a)$ function denotes set of all possible values of attribute $a\in Attr$. The intrinsic value is calculated as:

{\footnotesize
\begin{flalign*}
\begin{split}
IV(Ex,a) =  - & 
\sum \limits_{v \in {values(a)}}  \frac{| \big\{ x \in{Ex|value(x,a)=v} \big\} |}{|Ex|} \cdot \\
& \log_{2}\bigg( \frac{| \big\{ x \in{Ex|value(x,a)=v} \big\} |}{|Ex|} \bigg)
\end{split}
\end{flalign*}
}

\subsubsection{Hypotheses and variables}
We propose five null hypotheses:\\
\noindent$\bullet$~{\bf H10:} The 34 metrics have the same IGR.\\
\noindent$\bullet$~{\bf H20:} The five metric families have the same IGR.\\
\noindent$\bullet$~{\bf H30:} The IGR of R2RS metrics is not higher than the IGR of any other metric family.  \\
\noindent$\bullet$~{\bf H40:} The 18 R2RS metrics have the same IGR across different NLP techniques.  \\
\noindent$\bullet$~{\bf H50:} The 18 R2RS metrics have the same IGR across different distribution scores.  \\

The independent variables are the 34 metrics while the dependent variable is the IGR.
\subsubsection{Measurement and analysis procedure}
\label{sec:RQ1Measurement}
To determine the actual impact set of a requirement, we identify all the classes that are touched by the commits that include the requirement ID in their messages. This is based on the same assumption as when we compute the R2RS metric family (see Section \ref{sec:R2RS}), i.e. developers include the relevant requirement ID in their commit message when submitting code changes. For example, the impact set of a requirement named ACCUMULO-1009 includes the classes touched by any commit that includes the string \textit{[ACCUMULO-1009]} in the message. In practice, each issue is typically addressed by one commit \cite{raff}; however, sometimes multiple commits are associated with a single change.

The IGR of each of the 34 metrics is computed using WEKA \cite{weka} and is observed across each of the four projects.
The hypotheses of each research question is tested using the Kruskal-Wallis
test \cite{siegel1956nonparametric} which is similar to the more famous Anova test but does not have any assumptions about the distribution of the data. Because the non-parametric Kruskal-Wallis test is less powerful than the Anova test, it is more prone to not rejecting hypotheses when they actually need to be rejected, but on the other hand, when rejecting a hypothesis, it is more reliable than Anova. Moreover, the Kruskal-Wallis test is particularly recommended when the compared distributions are not independent. In our case, the distributions are computed over the same four projects and hence are not independent.
In this study we use a confidence level, i.e., alpha, of 5\% as standard in software engineering studies \cite{Wohlin:2012:ESE:2349018}.

\subsection{RQ2: How often are R2RS metrics selected by the classifiers to predict impacted classes?}
This research question aims to evaluate the effectiveness of the R2RS metrics for predicting impacted classes with automated metric selection techniques. 
Automated selection of metrics is usually performed by using a wrapper method. In the wrapper approach \cite{john1994irrelevant}, a metric subset selection algorithm is applied as a wrapper around the classifier. The algorithm conducts a search for a good subset of metrics using the specific classifier (as black box) and an objective function. In this paper we used the wrapper method that is provided as a default in WEKA and widely used: SubsetEvaluation \cite{Kohavi1997} in combination with BestFirst\footnote{http://weka.sourceforge.net/doc.stable/weka/\\attributeSelection/BestFirst.html}. In BestFirst the search is terminated when the accuracy is not improved by 0.1\% in the last 5 searches (explored combinations of metrics). A	thorough discussion of the subset evaluation is outside the scope of this paper and is already available in the literature \cite{john1994irrelevant, kohavi1997wrappers}.
The main advantage of a semi-exhaustive search is to provide a systematic approach to feasibly select the best combination of metrics. The main disadvantage is that the result is associated with a specific classifier(s).

% \subsubsection{Metrics selection}
% Intuitively, the best alternative to an unfeasible full-exhaustive search is to perform a semi-exhaustive search where specific algorithms and criteria are used to define and evaluate an incomplete set of combinations of metrics. The automated selection of metrics is usually performed by using a wrapper method. In the wrapper approach \cite{john1994irrelevant}, a metric subset selection algorithm is applied as a wrapper around the classifier. The algorithm conducts a search for a good subset of metrics using the specific classifier (as black box) and an objective function. In this paper we used the wrapper method that is provided as a default in WEKA and widely used: SubsetEvaluation \cite{Kohavi1997} in combination with BestFirst\footnote{http://weka.sourceforge.net/doc.stable/weka/attributeSelection/BestFirst.html}. In BestFirst the search is terminated when the accuracy is not improved by 0.1\% in the last 5 searches (explored combinations of metrics). A 	thorough discussion of the subset evaluation is outside the scope of this paper and is already available in the literature \cite{john1994irrelevant, kohavi1997wrappers}.
% The main advantage of a semi-exhaustive search is to provide systematic approach to feasibly select the best combination of metrics. The main disadvantage is that the result is associated with specific classifier(s).

\subsubsection{Diverse Classifiers}
\label{subsec:diverseClassifiers}
Because classification algorithms are highly sensitive to specific data characteristics, results from the experiments could be significantly biased if a single classifier were adopted.  We therefore compared results produced using five different classifiers -- all of which have been used in prior Software Engineering experiments of similar domains \cite{DBLP:journals/tse/MirakhorliC16, d2012evaluating, Kim:2011}. 

\noindent$\bullet$~{\bf Decision Tree}:
Decision Trees (DTs) are a supervised learning method used for classification and regression.  The goal is to create a model that predicts the type of a source file by learning simple decision conditions inferred from the words used in tactical files. In a decision tree, the internal nodes represent test conditions while leaf nodes represent categories.  In our case there are two categories of impacted and non-impacted classes. The attributes chosen to build the tree are based on information gain theory \cite{anyanwu2009comparative}, meaning that the internal nodes of the decision tree are constructed from the attributes that provide maximum information while the leaf nodes predict a specific category or class. To increase the accuracy of the decision tree, most algorithms employ some form of pruning to remove branches of the tree that are less useful in performing classifications \cite{anyanwu2009comparative}.  We use Weka's J48 decision tree because of its effectiveness in other software engineering studies \cite{Guo:2004}.\vspace{4pt}\\
\noindent$\bullet$~{\bf Random Forest}:
The Random Forest classifier generates a number of separate, randomized decision trees.  It has proven to be highly accurate and robust against noise \cite{breiman2001random}. However, because this classifier requires the building of many different decision trees, it can be extremely expensive to run on large projects.   \vspace{4pt}\\
\noindent$\bullet$~{\bf N\"aive Bayes}:
The N\"aive Bayes classifier is a probabilistic classifier based on Bayes theorem. It follows the assumption that the contribution of an individual feature toward deciding the probability of a particular class is independent of other features in that project instance \cite{mccallum1998comparison}. For example, in the case of detecting an impacted class, the contribution of the DIT feature is considered independent of the contribution of a Requirements-to-Code similarity feature. We included Naive Bayes because even though the assumption of independence does not hold in our feature set, the algorithm has been demonstrated to be effective in solving similar prediction problems \cite{Kim:2011, d2012evaluating, Guo:2004, Falessi2017}. \vspace{4pt}\\
\noindent$\bullet$~{\bf Logistic}:
The logistic classifier is built on a multinomial logistic regression model with a ridge estimator. Logistic regression model estimates the probabilities of the different possible outcomes of a categorically distributed dependent variable, given a set of independent variables. The estimation is performed through the logistic distribution function. The ridge estimator is used to improve the parameter estimation and to diminish future prediction error \cite{leCessie1992}.\vspace{4pt}\\
\noindent$\bullet$~{\bf Bagging}:
Bagging is an ensemble classifier. It trains individual classifiers on a random sampling of the original training set and then uses majority voting to attain the final results. Aggregation methods have been shown to be effective when used in conjunction with ``unstable'' learning algorithms such as decision trees where small changes in the training set result in large changes in predictions \cite{breiman1996bagging}.\\

\subsubsection{Hypotheses and variables}
We propose four additional null hypotheses:\\
\noindent$\bullet$~{\bf H60:} The 34 metrics are selected an equal proportion of times.\\
\noindent$\bullet$~{\bf H70:} The five metric families are selected an equal proportion of times.\\
\noindent$\bullet$~{\bf H80:} The 18 R2RS metrics are selected an equal proportion of times across different NLP techniques.  \\
\noindent$\bullet$~{\bf H90:} The 18 R2RS metrics are selected the same proportion of times across different distribution scores. \\ \\
The independent variables are the 34 metrics while the dependent variable is the proportion of times a metric is selected among different classifiers and projects.

\subsubsection{Analysis Procedure}
The proportion of times a metric is selected has been computed by using standard WEKA parameters (i.e., \textit{weka.attributeSelection.WrapperSubsetEval -B} and \textit{weka.attributeSelection.BestFirst -D 1 -N 5}).

The hypotheses of these research questions are tested using the Fisher's exact test \cite{Fisher22}. This is a non-parametric test which is suitable when the dependent variable is categorical (i.e., whether a metric is selected or not).

\subsection{RQ3: What is the accuracy of single metrics families in predicting impacted classes?}
This research question aims to evaluate the prediction accuracy (i.e., F1) achieved by using one family metric at a time.
\subsubsection{Hypotheses and variables}
We propose one null hypotheses: \\
\textbf{H100:} The five metric families have the same prediction accuracy.\\ \\
The independent variable is the single family of metrics used as input of the prediction. The dependent variable is F1. We select F1 over other F\textit{x} values, because both recall and precision are important for impact analysis tasks.  Furthermore, F1 has previously been widely adopted in similar studies \cite{kagdi2013integrating}. This is defined as a combination of the following accuracy metrics. 

\noindent{\bf True Positives (TP)}: A class is correctly predicted to be impacted by a new requirement.\vspace{4pt}\newline
\noindent{\bf False Positives (FP)}: A class is predicted to be impacted by a new requirement, even though it is not.\vspace{4pt}\newline
\noindent{\bf True Negatives (TN)}:  A class is correctly predicted to not be  impacted by a new requirement.\vspace{4pt}\newline
\noindent{\bf False Negatives (FN)}: A class is predicted to not be impacted, even though it is actually impacted. \vspace{4pt}\newline
\noindent{\bf Precision}:~represents the percentage of times a class that is predicted to change actually changes. It is defined as:
\begin{equation}
\label{eq:Precision}
\textit{Precision} =  \frac{\textit{TP}}{\textit{TP + FP}}
\end{equation}

\noindent{\bf Recall}: represents the percentage of times a class that changes was predicted to change. It is defined as: 
\begin{equation}
\label{eq:Recall}
Recall =  \frac{TP}{TP + FN}
\end{equation}

\noindent{\bf  F1}: in general, F\textit{x} measures the classification accuracy with respect to a user who attaches \textit{x} times as much importance to Recall as Precision. F1 assumes that Precision and Recall have the same importance and is defined as: 
\begin{equation}
\label{eq:Fx}
F1 =  2 * \frac{Precision * Recall}{Precision + Recall}
\end{equation}

Note that individual Recall and Precision results are reported in Table \ref{tab:pre_rec_rq3} in the Appendix.  Further, we do not report results using other common measures such as Receiver Operating Characteristic (ROC) or Mean Average Precision (MAP) because our classifier is binary in nature and therefore neither delivers a range of Recall/Precision values nor a ranking of the predicted classes \cite{DBLP:conf/icse/ShinHC15}. 

\subsubsection{Measurement and analysis procedure}
To train each individual classifier, we created a time-ordered 80\%-20\% split of the data.  We were unable to use a standard N-Fold cross-validation approach \cite{arlot2010survey,witten2005data, Kim:2011} because selecting any fold other than the final fold for testing purposes would violate the time-sequencing of R2RS and temporal history features.  This would have led to the improper scenario in which future knowledge was used to predict past-events.  We therefore first split the data into a training set containing 80\% of the data and a testing set containing the remaining 20\%.  

Our data was severely imbalanced with respect to the class \emph{requirement impacts class (yes/no)} for three of the four projects. For example, in the case of the Accumulo project containing 85,955 candidate requirements-to-class pairs, there were only 3,412 (or $\approx$ 3.97\%) represented cases in which the requirement had impacted the class, while the remaining 82,543 (or $\approx$ 96.03\%) were not impacted. Training a classifier on imbalanced data is highly problematic \cite{chawla2005data} because there is a tendency for the majority of instances to be classified into the dominant class.  In our case, classifying all instances as \emph{not impacted} would produce extremely high overall accuracy but would completely fail to achieve the goal of identifying impacted classes.  We informally evaluated two common approaches for handling unbalanced data, namely \emph{over-sampling} the minority case and \emph{under-sampling} the majority case.  Over-sampling produced very low accuracy in predictions, likely because  the huge imbalance caused excessive duplication of each impacted requirement-class pair, leading to over-training.  We therefore adopted the under-sampling method in which all impacted classes were retained in the training set, while an equal number of non-impacted classes were randomly selected.  The test set was not modified.   Due to the under-sampling, many examples of non-impacted data instances were discarded in each evaluation.  We therefore repeated the under-sampling 20 times for each experiment and report results for each sample. Results were then averaged among the 20 samples. The standard deviation returned an average of 0.027 for F1 (our  dependent variable) across classifiers and data sets.  This suggests that the number of samples was sufficient for avoiding bias in the random sampling procedure. 
The classifiers have been used with standard parameters as reported in Table \ref{tab:WekaParams}.
The hypothesis of these research questions are tested using the Kruskal-Wallis
test \cite{siegel1956nonparametric} for the same reasons as described in Section \ref{sec:RQ1Measurement}.

\begin{table}
%\begin{center}
 \caption{Default classifiers parameters in WEKA}
 \label{tab:WekaParams}
 \begin{tabular}{|L{2.4cm}|L{5.2cm}|} 
 \hline
 Classifier & Parameters\\ [0.5ex] 
 \hline\hline
 Bagging & -P 100 -S 1 -l 10 -W weka.classifiers.trees.REPTree\newline-M 2 -V 0.001 -N 3 -S 1 -L -1 \\ 
 \hline
  J48 & -C 0.25 -M 2 \\
 \hline
 Logistic & -R 1.0E-8 -M -1 \\
 \hline
 Naive Bayes & Default classifier- \\
 \hline
 Random Forest & -l 100 -K 0 -S 1 \\
 \hline

\end{tabular}
%\end{center}
\end{table}

\subsection{RQ4: How much does using  R2RS improve the accuracy in predicting impacted classes?}
This research question aims to evaluate the practical gain in prediction accuracy (i.e., F1) achieved by using R2RS metrics as input, in addition to the other 16 metrics, to prediction models.
\subsubsection{Hypotheses and variables}
We propose one null hypotheses: \\
\textbf{H110:} Leveraging R2RS metrics does not increase the accuracy in predicting impacted classes.\\
The independent variable is whether R2RS metrics are used in the prediction of impacted classes. The depended variable is F1.
\subsubsection{Measurement and analysis procedure}
The measurement and analysis procedure coincides with the one we adopted for RQ3.

\subsection{Experimental Objects}
\label{subsec:exp_obj}
The efficacy of the five metric families was evaluated against four large open-source Java projects.  The number of projects was influenced by resource availability and by the high number and extreme diversity of the metrics we needed to measure for each revision of the project.  We used the Apache repository\footnote{https://people.apache.org/phonebook.html} as the source of projects, considering projects that: (1) were written in Java, (2) stored tickets in JIRA, and (3) used Git version control. We focused on Apache projects, rather than random GitHub projects, to avoid using toy projects \cite{munaiah2016curating}.
 For selecting projects we used the guidelines provided by Nagappan et al. \cite{nagappan2013diversity} including the fact that a higher number of projects is not necessarily better and that it is important to have a desired level of diversity among projects. Moreover, generally speaking, the more varied the data are the more accurate and robust the classifier can be trained through these data \cite{Domingos:2012}. Thus, we prioritized projects with a higher number of commits, larger numbers of requirements, and higher proportions of linked requirements to commits. We first selected the top eight Apache projects with the highest proportion of linked commits to requirements and then we ranked them  according to the following formula: \emph{Number of Requirements ~ $\times$~ Percentage of Linked Commits ~ + ~0.3~$\times$~ Number of Commits}. The rationale of this equation is to make a trade-off among quality (percentage of linked commits) and representativeness (size in terms of requirements and commits) of the project. As a result of this process, four projects were selected. We measured projects data in April 2016. We note that our project selection approach is replicable \cite{Falessi:STRESS}.
In all four projects Jira issues were tagged either as \emph{bug fixes} or as \emph{new features}. A new feature request submitted to the Jira issue tracker represents a user's request for new or improved functionality. While they are specified less formally than a traditional requirement, they serve the same purpose of describing new, desired functionality. In our study, we focused only on predicting the impact of new features.

The four projects included in our study are:

\noindent$\bullet$~{\bf Accumulo} is a distributed key-value database system that bears some similarity to Amazon's DynamoDB \cite{apache-accumulo}. It was launched in 2008 by the US National Security Agency. It has an extremely large code-base, with new requirements implemented in each release.  We retrieved 145 requests for new features extracted from 7,889 commits between $10/6/11$ and $12/19/14$.   The size of the code-base grew as the project evolved; however there was an average of 593 classes at the time each metric was computed. Only 3.97\% of candidate requirements-class pairs represented positive impacts. Sample requirements are shown in Table \ref{tab:Requirements} and were previously discussed in Section \ref{sec:metrics}. \vspace{4pt}\\
\noindent$\bullet$~{\bf Ignite} is a high-performance, integrated, and distributed in-memory platform for computing and real-time transacting on large-scale data sets.  Process is much faster than is possible with traditional disk-based or flash technologies \cite{apache-ignite}. Our data-set includes 41 feature requests extracted from 16,571 commits over the period of $11/14/14$ to $01/11/16$, and an average code base of 668 classes; here, 56.85\% of candidate requirements-class pairs represented actual impact points, meaning that this project was only balanced with respect to impacted and non-impacted pairs.  Sample requirements include:\\
\noindent -\emph{Instead of ordering messages on receiver side, we can do ordering on sender side.}\\ 
\noindent -\emph{Batch mode for {\bf visorcmd} allows to read commands from test file (one command per line), perform them, send output to STDOUT/STDERR and exit if end of text file has been reached}. \vspace{4pt}\\ 
\noindent$\bullet$~{\bf Isis} is a framework developed in Java and designed to support rapid web development. It contains a ready-made UI component as well as RESTful services \cite{apache-isis}. We retrieved 252 requests for new features over the period of 10/4/10 to 02/16/16, and an average code base consisting of 2,424 classes.  Only 1.94\% of candidate requirements-class pairs represented positive impacts.  Sample requirements include:\\
\noindent -\emph{To safe (sic) the users some time in navigation the application name (or logo) above the sign up and password reset forms could be a link to the sign in page.}\\
\noindent -\emph{Isis standardized on using JMock a while back, however as a library it doesn't seem to be keeping up...Candidates to replace it are mockito and jmockit.} \vspace{4pt}\\
\noindent$\bullet$~{\bf Tika} is an open-source development project dedicated to parsing a variety of different file-types and extracting their metadata \cite{apache-tika}. It has been in development since June of 2007, with its first major release in December of the same year. The Tika project is our smallest project, consisting of 49 requirements, an average of 72 classes, and requirement-class impacts of 7.09\%.  Data was collected for the period of $10/11/07$ to $03/12/14$. A sample requirement is:\\
\noindent -\emph{Currently, the text parser implementation uses the default encoding of the Java runtime when instantiating a Reader for the passed input stream.  We need to support other encodings as well. It would be helpful to support the specification of an encoding in the parse method. Ideally, Tika would also provide the ability to determine the encoding automatically based on the data stream.}  
\vspace{4pt}

The detailed characteristics of each project are summarized in Table \ref{tab:projects}.  
The ``No. of Releases'' row represents the total number of releases we analyzed for a specific project.
The ``No. of Commits Linked to Code'' row represents the total number of commits that were associated with changes to source code. The ``No. of Requirements'' represents the number of feature requests in the project. The ``Average No. of Classes''represents the average number of classes that existed at the time each requirement started to be implemented.  This is an average because the total number of classes changed over the timeline of the project.  The ``No. of Candidate Req.-Class Pairs'' represents the number of combinations between all requirements and all classes. The ``\% of Positive Instances'' represents the number of classes touched by a requirement divided by the number of candidate requirements-class pairs. Finally, the two date rows represent the dates of the first and last commit for each project. 

\subsection{Replication}
For replication purposes, we provide (1) the source files of the application developed to collect the 34 metrics\footnote{www.falessi.com/CCP/CCPsource.zip}, (2) the resulting arff files\footnote{www.falessi.com/CCP/CCParff.zip}, and, (3) the WEKA output\footnote{www.falessi.com/CCP/CCPWEKAresults.zip} of RQ2 and RQ3. This WEKA output also reports the  WEKA parameters used to investigate RQ2. Finally, to further support repeatability of our approach, Table \ref{tab:WekaParams} reports the parameters used to train each of the classifiers to investigate RQ3.

%% file: Results.tex
\section{Results and Discussion}
\label{sec:Results}
We now discuss the findings with respect to each of the research questions. 
\subsection{RQ1: Do  R2RS  metrics  provide  important information to predict impacted classes?}

\begin{table}[!t]
\setlength{\tabcolsep}{0pt}
\caption{Statistics for the four projects.  }
\label{tab:projects}
\begin{tabular}{|L{3cm}|L{1.4cm}|L{1.3cm}|L{1.3cm}|L{1.3cm}|}
 \hline
\textbf{Project Characteristics} & \textbf{Accumulo} & \textbf{Ignite} & \textbf{Isis} & \textbf{Tika} \\  \hline
No. of Releases    & 8              & 23           & 20          & 27          \\  \hline
No. of Commits Linked to Code    & 7,889              & 16,571           & 6,521          & 2,781          \\  \hline
No. of Requirements                & 145               & 41              & 252           & 49            \\  \hline
Average No. of Classes               & 593             & 668            &  2,424         & 72           \\  \hline
No. of Candidate Req--Class Pairs          & 85,955             & 27,391           & 610,833        & 3,508          \\  \hline
\% of Positive Instances  & 3.97\%             & 56.84\%           & 1.94\%        & 7.09\%          \\  \hline
Earliest Create Date             & 10/6/11           & 11/14/14        & 10/4/10       & 10/11/07      \\  \hline
Latest Commit Date                  & 12/19/14          & 01/11/16         & 02/16/16       & 03/12/14 \\ \hline    
\end{tabular}
\vspace{-16pt}
\end{table}

In order to investigate how the IGR of metrics  varies among projects, we computed the rank of each individual metric according to IGR for each project; we then computed the median rank among projects.  Table \ref{tab:IGRranks} depicts the results. We make the following observations:\\
$\bullet$~  The IGR varies across metrics. In fact, the Kruskal-Wallis test comparing the IGR of different metrics provides a p-value of 0.0004. Therefore, we can reject H10 and can claim that there is a statistically significant difference in the IGR provided by different metrics. \\
$\bullet$~ R2RS metrics have a median IGR higher than all metrics of all other families other than TLCC. \\
$\bullet$~TLCC\_Lin is the metric with the highest median rank, i.e., it exhibits the highest IGR among different projects.\\
$\bullet$~The top ranked metric changes among projects. Specifically, it is R2RS\_BC\_Max in Ignite, and TLCC\_Lin in Accumulo, Isis, and Tika. Note that the R2RS\_BC\_Max, which is the best in Ignite, is the worst in the remaining three projects.\\
$\bullet$~CKJM\_NOC is the lowest ranked metric.\\
$\bullet$~The top ranked R2RS metric is VSM\_Max. The VSM method with the Maximum distribution score is the most effective for measuring the Requirement to Requirements Set similarity for the purpose of impact analysis. It implies that the key words weighted by tf-idf scheme can potentially serve as strong indicators for determining the impact of future changes.\\
$\bullet$~The lowest ranked R2RS metrics are the ones associated with the Greedy Comparator (GC). The terms might be incorrectly paired during the term-to-term semantic similarity calculation due to its greedy pairing mechanism. Therefore, the actual semantic similarity between the requirement and the requirement set might not be adequately captured. \\
$\bullet$ All of the 18 R2RS metrics have a median rank higher than well established change proneness metrics such as size SQ\_NCLOC and smells SQ\_Viol \cite{khomh2009exploratory, arvanitou2017method, zhou2009examining}. This means that focused information, about the specific requirement to implement, outperforms coarse grained information about the source code such as size and smells.\\

% \begin{table}[!b]
% \label{tab:IGRranks}
% \caption{IGR of metrics family across four projects.}
% \includegraphics[width=\columnwidth]{IGRranks.pdf}

% \end{table}

\begin{table*}[t]
\centering
\caption{IGR of metrics in four projects.}
\label{tab:IGRranks}
\begin{tabular}{|l|c|c|c|c|c|c|c|c|c|c|}
\hline
\multicolumn{1}{|c|}{\multirow{2}{*}{Metric}} & \multicolumn{2}{c|}{ACCUMULO} & \multicolumn{2}{c|}{IGNITE} & \multicolumn{2}{c|}{ISIS} & \multicolumn{2}{c|}{TIKA} & \multirow{2}{*}{Median Rank} \\ \cline{2-8}
\multicolumn{1}{|c|}{}                        & IGR            & Rank         & IGR           & Rank        & IGR          & Rank       & IGR          & Rank       &                                                            \\ \hline
TLCC\_Lin                                     & 0.158          & 1            & 0.063         & 20          & 0.261        & 1          & 0.425        & 1          & 1                                                         \\ \hline
TLCC\_Log                                     & 0.098          & 2            & 0.107         & 4           & 0.173        & 2          & 0.067        & 2          & 2                                                         \\ \hline
R2RS\_VSM\_Max                                & 0.011          & 11           & 0.093         & 5           & 0.004        & 5          & 0.044        & 3          & 5                                                         \\ \hline
R2RS\_JSD\_Max                                & 0.01           & 13           & 0.085         & 9           & 0.004        & 4          & 0.041        & 4          & 7                                                        \\ \hline
R2RS\_JSD\_Top5                               & 0.01           & 14           & 0.091         & 6           & 0.004        & 7          & 0.037        & 6          & 7                                                        \\ \hline
R2RS\_CMC\_Av                                 & 0.033          & 3            & 0.07          & 19          & 0.003        & 8          & 0.035        & 7          & 8                                                       \\ \hline
R2RS\_CMC\_Max                                & 0.017          & 4            & 0.09          & 7           & 0.003        & 15         & 0.035        & 9          & 8                                                         \\ \hline
R2RS\_CMC\_Top5                               & 0.016          & 5            & 0.079         & 16          & 0.003        & 11         & 0.035        & 8          & 10                                                      \\ \hline
R2RS\_JSD\_Av                                 & 0.009          & 16           & 0.082         & 12          & 0.004        & 6          & 0.032        & 10         & 11                                                       \\ \hline
R2RS\_OPC\_Max                                & 0.016          & 6            & 0.085         & 8           & 0.003        & 17         & 0.018        & 14         & 11                                                        \\ \hline
R2RS\_VSM\_Top5                               & 0.01           & 12           & 0.082         & 13          & 0.003        & 14         & 0.039        & 5          & 13                                                        \\ \hline
R2RS\_OPC\_Top5                               & 0.013          & 8            & 0.08          & 15          & 0.003        & 13         & 0.018        & 12         & 13                                                       \\ \hline
R2C\_JSD                                      & 0.012          & 9            & 0.009         & 25          & 0.003        & 9          & 0.01         & 18         & 14                                                        \\ \hline
R2RS\_VSM\_Av                                 & 0.01           & 15           & 0.075         & 17          & 0.003        & 12         & 0.032        & 11         & 14                                                       \\ \hline
TLCC\_SCP                                     & 0.013          & 7            & 0.057         & 21          & 0.009        & 3          & 0            & 28         & 14                                                        \\ \hline
R2RS\_OPC\_Av                                 & 0.012          & 10           & 0.074         & 18          & 0.003        & 16         & 0.018        & 13         & 15                                                       \\ \hline
R2RS\_BC\_Top5                               & 0.006          & 22           & 0.118         & 2           & 0.001        & 18         & 0.011        & 17         & 18                                                       \\ \hline
R2C\_VSM                                      & 0.007          & 18           & 0.005         & 31          & 0.003        & 10         & 0.009        & 19         & 19                                                     \\ \hline
R2RS\_BC\_Av                                 & 0.006          & 21           & 0.116         & 3           & 0.001        & 23         & 0.011        & 16         & 19                                                       \\ \hline
R2RS\_GC\_Av                                 & 0.006          & 20           & 0.081         & 14          & 0.001        & 21         & 0            & 33         & 21                                                      \\ \hline
CKJM\_RFC                                     & 0.008          & 17           & 0.009         & 24          & 0            & 26         & 0.006        & 21         & 23                                                      \\ \hline
R2RS\_BC\_Max                                & 0              & 33           & 0.136         & 1           & 0.001        & 20         & 0            & 25         & 23                                                      \\ \hline
R2RS\_GC\_Top5                               & 0.005          & 23           & 0.082         & 11          & 0.001        & 22         & 0            & 31         & 23                                                       \\ \hline
CKJM\_CBO                                     & 0.003          & 25           & 0.012         & 22          & 0            & 25         & 0.012        & 15         & 24                                                       \\ \hline
CKJM\_WMC                                     & 0.007          & 19           & 0.009         & 27          & 0            & 27         & 0.006        & 20         & 24                                                       \\ \hline
R2RS\_GC\_Max                                & 0.002          & 30           & 0.083         & 10          & 0.001        & 19         & 0            & 32         & 25                                                       \\ \hline
SQ\_NCLOC                                     & 0.002          & 26           & 0.009         & 26          & 0            & 28         & 0.004        & 23         & 26                                                       \\ \hline
CKJM\_NPM                                     & 0.004          & 24           & 0.006         & 30          & 0            & 30         & 0.005        & 22         & 27                                                       \\ \hline
CKJM\_LCOM                                    & 0.002          & 29           & 0.008         & 29          & 0            & 29         & 0            & 24         & 29                                                       \\ \hline
SQ\_Com                                       & 0.002          & 28           & 0.01          & 23          & 0            & 31         & 0            & 30         & 29                                                      \\ \hline
SQ\_Viol                                      & 0.002          & 27           & 0.008         & 28          & 0            & 32         & 0            & 29         & 29                                                       \\ \hline
CKJM\_DIT                                     & 0.001          & 31           & 0.003         & 32          & 0            & 33         & 0            & 26         & 32                                                      \\ \hline
CKJM\_CA                                      & 0.001          & 32           & 0.002         & 33          & 0            & 34         & 0            & 27         & 33                                                       \\ \hline
CKJM\_NOC                                     & 0              & 34           & 0             & 34          & 0.001        & 24         & 0            & 34         & 34                                                      \\ \hline
\end{tabular}
\end{table*}

\begin{figure}[!b]
\includegraphics[width=\columnwidth]{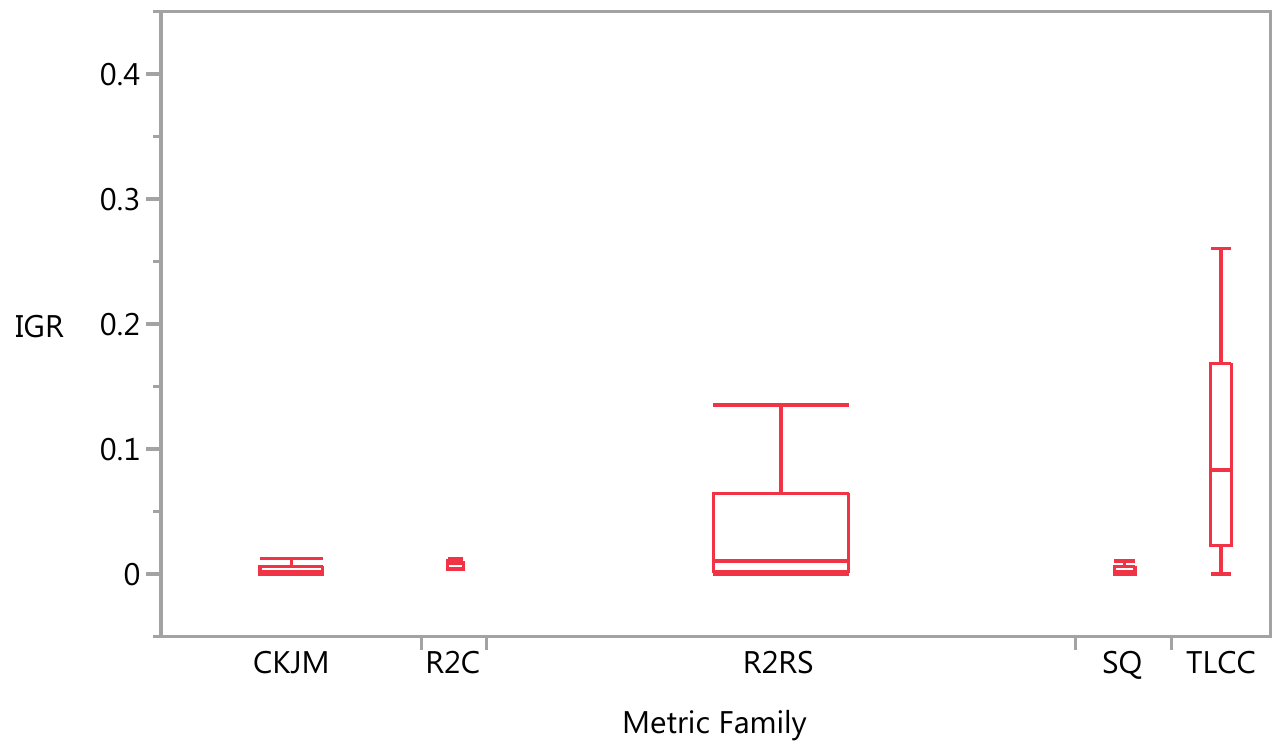}
\caption{Distributions of IGR of metrics family across the four projects.}
 \label{fig:RQ1byFamily}
\end{figure}

We computed the average IGR among metrics of the same family and it returned 0.003 for SQ, 0.003 for CKJM, 0.007 for R2C, 0.031 for R2RS, and 0.119 for TLCC.
Figure \ref{fig:RQ1byFamily} reports the distribution of IGR values among metrics of the same family.
We make the following observations:\\
$\bullet$~ The IGR varies across metrics families. In fact, the Kruskal-Wallis test comparing the IGR of different metrics provides a p-value of 0.0001. Therefore we can reject H20 and can claim that there is a statistically significant difference in the IGR provided by different metric families. \\
$\bullet$~ The median IGR of TLCC metrics is much higher than the median of any other metric family.  \\
$\bullet$~ The Kruskal-Wallis test comparing the IGR of R2RS versus other metric families provides a p-value of 0.0001 versus CKJM, of 0.265 versus R2C, of 0.0006 versus SQ, and of 0.004 versus TLCC. Therefore \noindent \emph{we can reject H30 in two out of four cases} and we can claim that R2RS metrics have an IGR that is statistically higher than both CKJM and SQ. We note that R2RS metrics have an IGR that is statistically lower than TLCC.   

\begin{figure}[!t]
\includegraphics[width=\columnwidth]{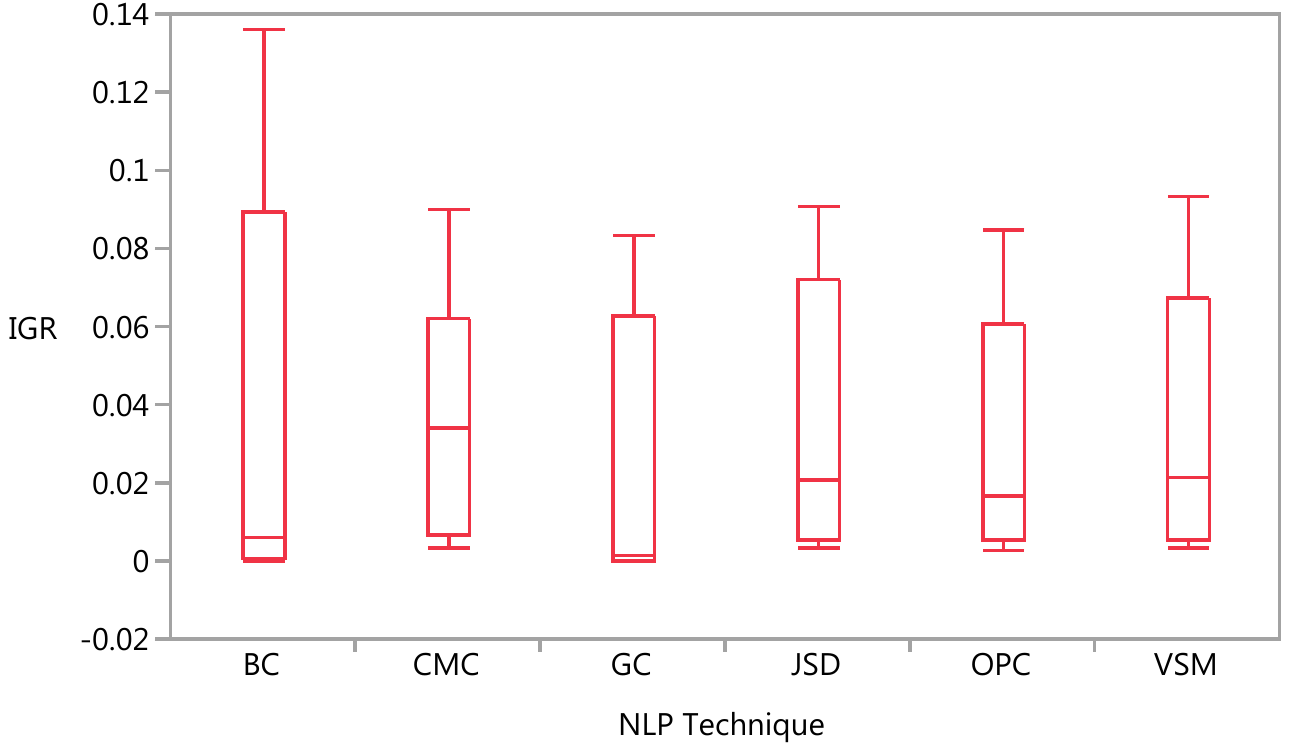}
\caption{Distributions of IGR among NLP techniques.}
 \label{fig:RQ1byTechnique}
\end{figure}

We computed the average IGR among metrics of the same R2RS NLP technique and it returned 0.035 for CMC, 0.033 for JSD, 0.034 for BC, 0.022 for GC, 0.034 for VSM, and 0.029 for OPC. Figure \ref{fig:RQ1byTechnique} reports the distributions of IGR among NLP techniques. The Kruskal-Wallis test comparing the IGR of NLP techniques resulted higher than alpha. Therefore, \noindent \emph{we cannot reject H40}; thus, we cannot claim IGR varies among the investigated NLP techniques.

\begin{figure}[!t]
\includegraphics[width=\columnwidth]{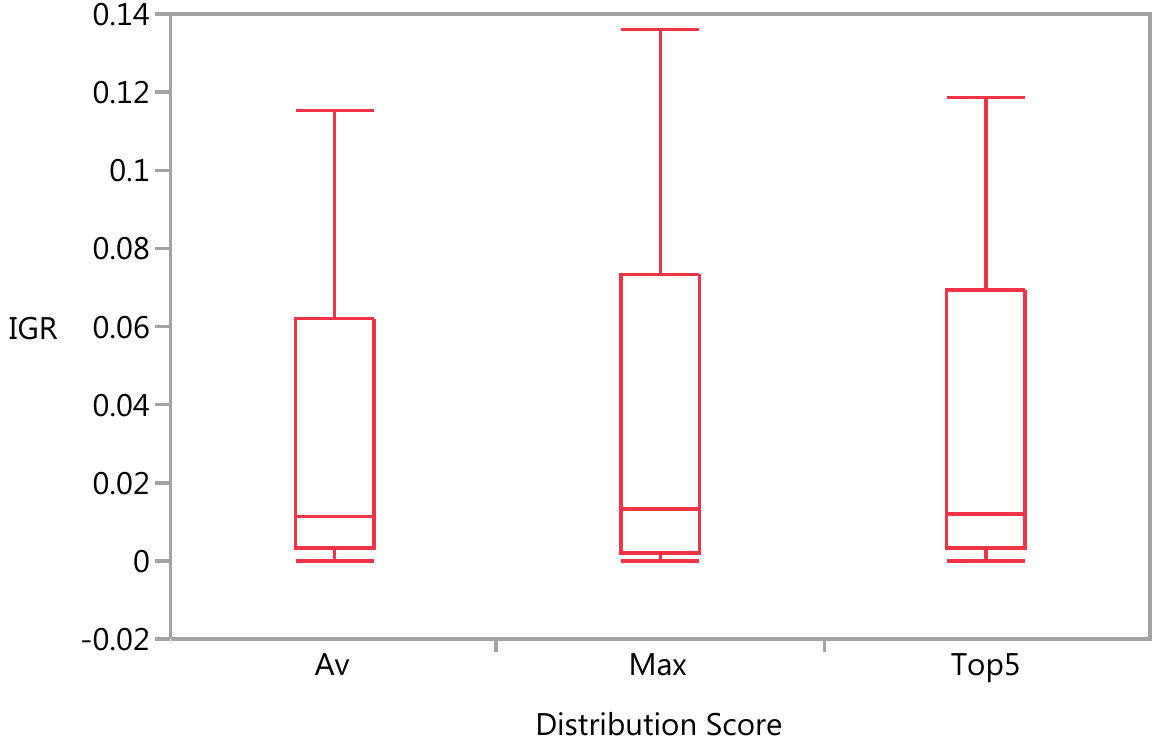}
\caption{Distributions of IGR among distribution scores.}
 \label{fig:RQ1byApproach}
\end{figure}

We computed the average selection proportions among metrics of the same R2RS distribution score and it returned 0.030\% for Avg, 0.035 for Max, and 0.031 for Top5. Figure \ref{fig:RQ1byApproach}  reports the distributions of IGR among distribution scores. The Kruskal-Wallis test comparing the IGR of distribution scores was higher than alpha. Therefore, we \noindent \emph{can neither reject H50} nor claim IGR varies across distribution scores.

\begin{center}
\fbox{
\begin{minipage}[t]{.95\linewidth}
{\bf RQ1 Summary:}
The amount of information provided to predict impacted classes significantly varies among metrics and metrics families. The R2RS metrics family provides an amount of information that is significantly higher than CKJM, R2C and SQ, but lower than TLCC.
\end{minipage}
}
\end{center}

\subsection{RQ2: Are R2RS metrics selected to predict impacted classes?}

\begin{figure}[!b]
\includegraphics[width=\columnwidth]{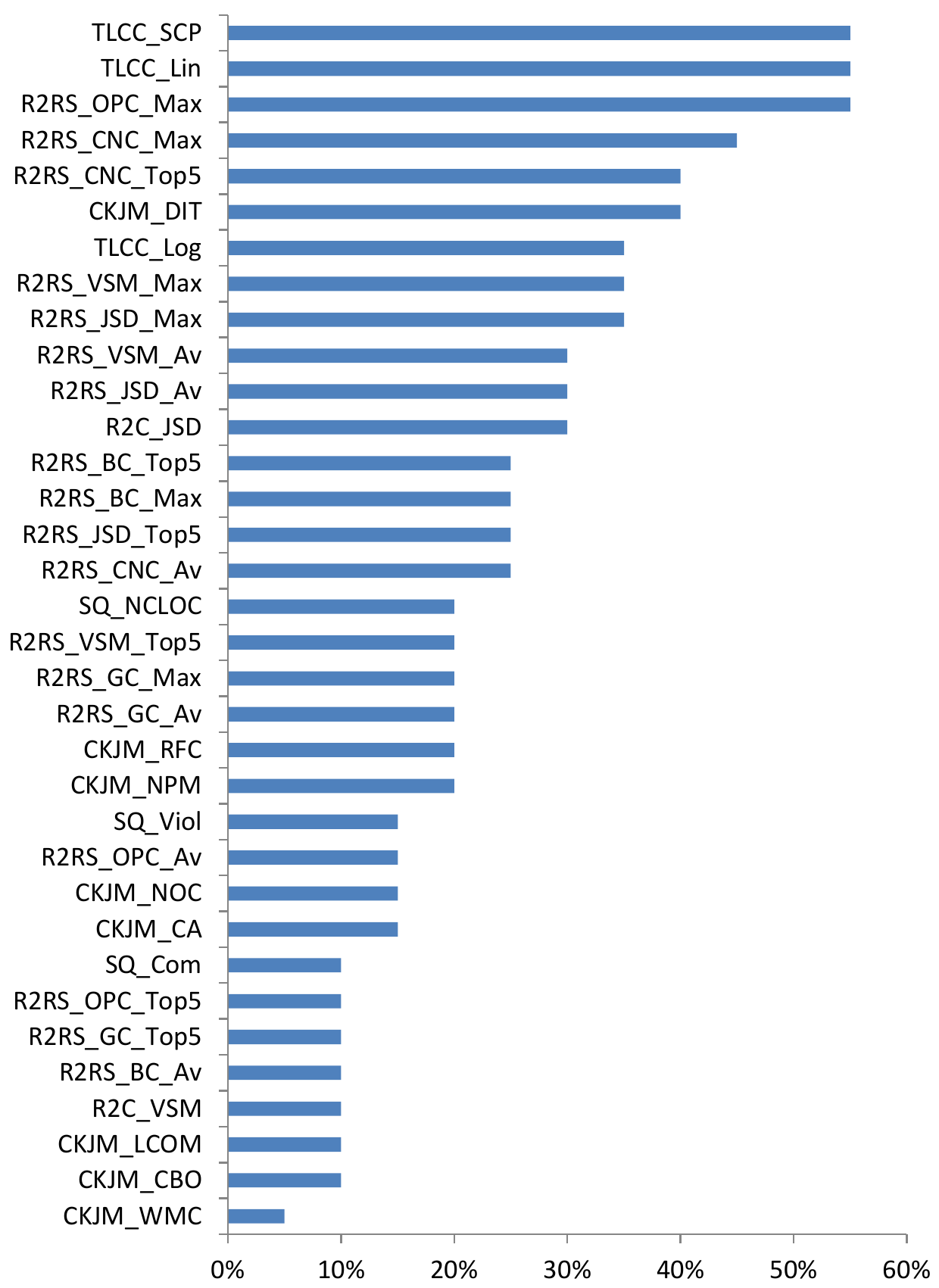}
\caption{Proportions of times a metric has been selected to maximize accuracy among five classifiers and four projects.}
 \label{fig:SelectionFrequency}
\end{figure}

Table \ref{tab:MetricsSelectionLarge} and Table \ref{tab:MetricsSelectionCompact} in the Appendix report, in two different formats, the specific metrics selected by a specific classifier in a specific project. Figure \ref{fig:SelectionFrequency}, by summarizing the content of  Table \ref{tab:MetricsSelectionLarge} and Table \ref{tab:MetricsSelectionCompact}, reports the proportions of times a metric has been selected from among the five classifiers and the four projects. We make the following observations:\\
$\bullet$~ Only three out of 34 metrics have been selected in more than half of the cases. These three metrics are R2RS\_OPC\_Max, TLCC\_Lin, TLCC\_SCP and they all have been selected in 11 out of 20 cases. \\
$\bullet$~ TLCC\_Lin resulted among the top ranked metrics in both IGR and selection proportions. 
\\
$\bullet$~ The least selected metric is CKJM\_WMC. Moreover, the least three selected metrics are all of the CKJM family.\\
$\bullet$~Based on the fact that the most selected metric was selected in just over half of the cases, and no metric was never selected, we conclude that the importance of a metric depends on the project and the classifier adopted.\\
$\bullet$~ 11 out of 18 R2RS metrics have a selection proportion higher than well established change proneness metrics such as size (SQ\_NCLOC) and smells (SQ\_Viol) \cite{khomh2009exploratory, arvanitou2017method, zhou2009examining}. Again, this means that focused information, about the specific requirement to implement, outperforms coarse grained information of the source code such as size and smells.\\
$\bullet$~  The Spearman correlation between the IGR rank and the selection proportion is of 0.609 (P-value = 0.0001) which suggests a high but not perfect agreement among the two ranks. \\
$\bullet$~ The Fisher's exact test comparing the proportions of selections among metrics returns a p-value  of 0.0001. Therefore, \noindent \emph{we can reject H60} and can claim that there is a difference in the proportion of times each metric is selected. 

\begin{figure}[b]
\includegraphics[width=\columnwidth]{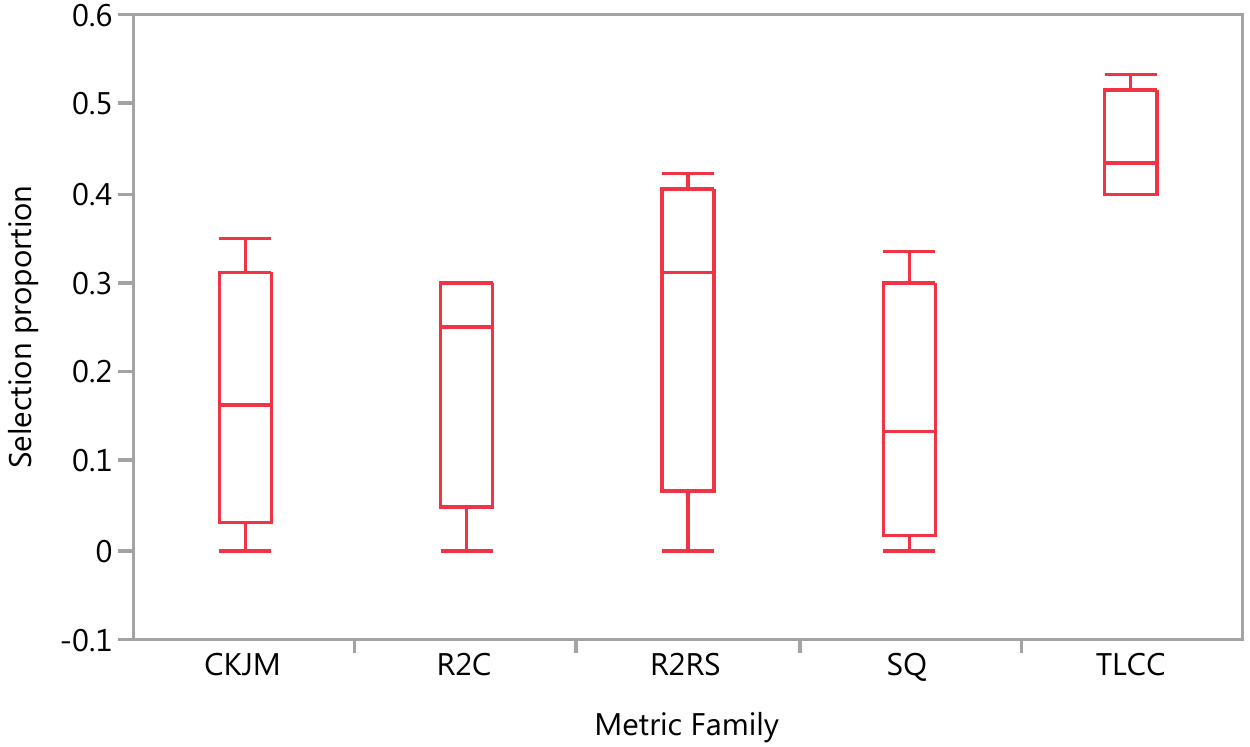}
\caption{Distributions of selection proportions of different metric families across the four projects.}
 \label{fig:RQ2family}
\end{figure}

We computed the average selection proportions among metrics of the same family and it returned 15\% for SQ, 17\% for CKJM, 20\% for R2C, 26\% for R2RS, and 45\% for TLCC. Figure \ref{fig:RQ2family} reports the distribution of selection proportion of different metric families across the four projects. The Fisher's exact test comparing the proportions of selections among metrics of different families returned a p-value  of 0.0003. Therefore, \noindent \emph{we can reject H70} and we can claim that the proportion of times a metric is selected differs among metrics families.

\begin{figure}[b]
\includegraphics[width=\columnwidth]{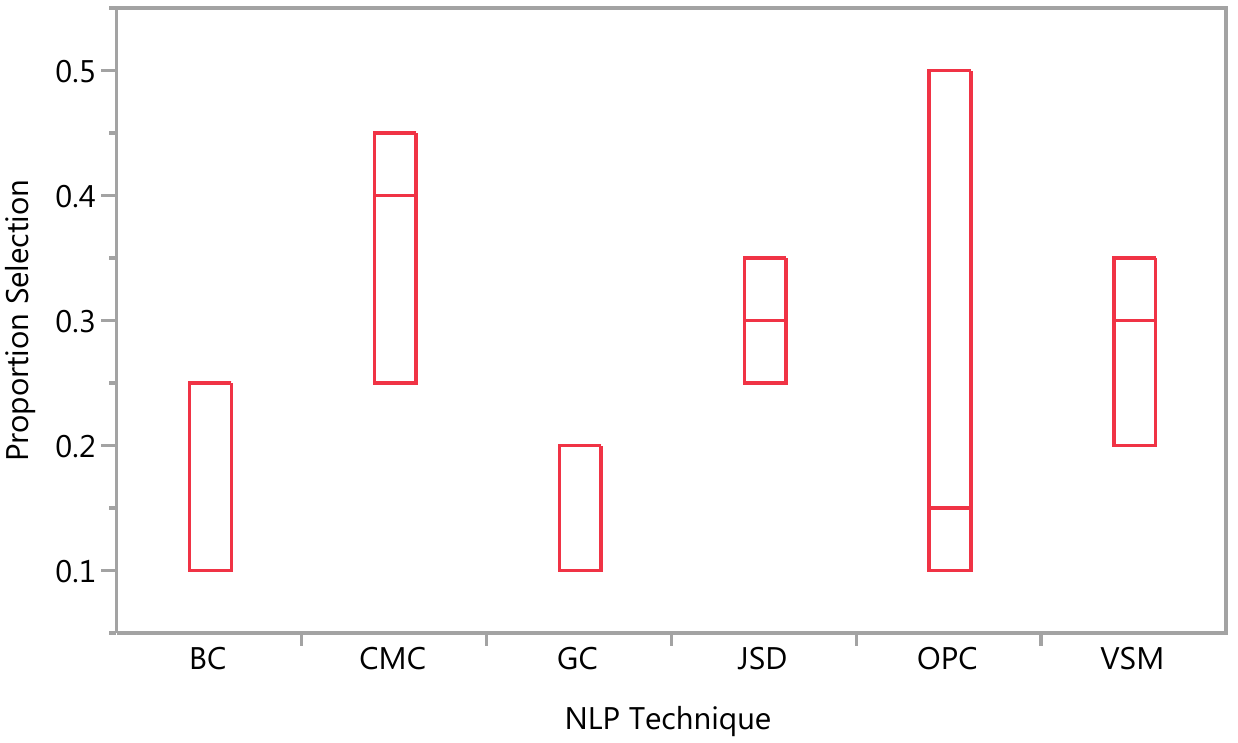}
\caption{Distributions of selection proportions of different R2RS NLP techniques across the four projects.}
 \label{fig:RQ2technique}
\end{figure}

We computed the average selection proportions among metrics of the same R2RS NLP technique and it returned 37\% for CMC, 30\% for JSD, 20\% for BC, 17\% for GC, 28\% for VSM, and 45\% OPC. Figure \ref{fig:RQ2technique} reports the distribution of selection proportion of different R2RS NLP techniques across the four projects. The Fisher's exact test comparing the selection proportions among R23RS NLP techniques returned a p-value  of 0.1545. Therefore, \noindent \emph{we cannot reject H80} and we cannot claim that the proportions of times a R2RS metric is selected differs among NLP techniques. 

We computed the average selection proportions among metrics of the same R2RS distribution score and it resulted in 22\% for Av, 35\% for Max, and 22\% Top5. Figure \ref{fig:RQ2technique} reports the distribution of selection proportion of different R2RS distribution scores across the four projects. The Fisher's exact test comparing the proportions of selections among metrics of different families results a p-value  of 0.0301. Therefore, \noindent \emph{we can reject H90} and we can claim that the proportions of times a R2RS metric is selected differs among distribution scores. We note that the distribution score with the highest selection proportion (i.e., Max) coincides to the NLP technique with the highest IGR. Thus, it is advised to use the maximum similarity score over the average or the top five similarity scores.

\begin{figure}[b]
\includegraphics[width=\columnwidth]{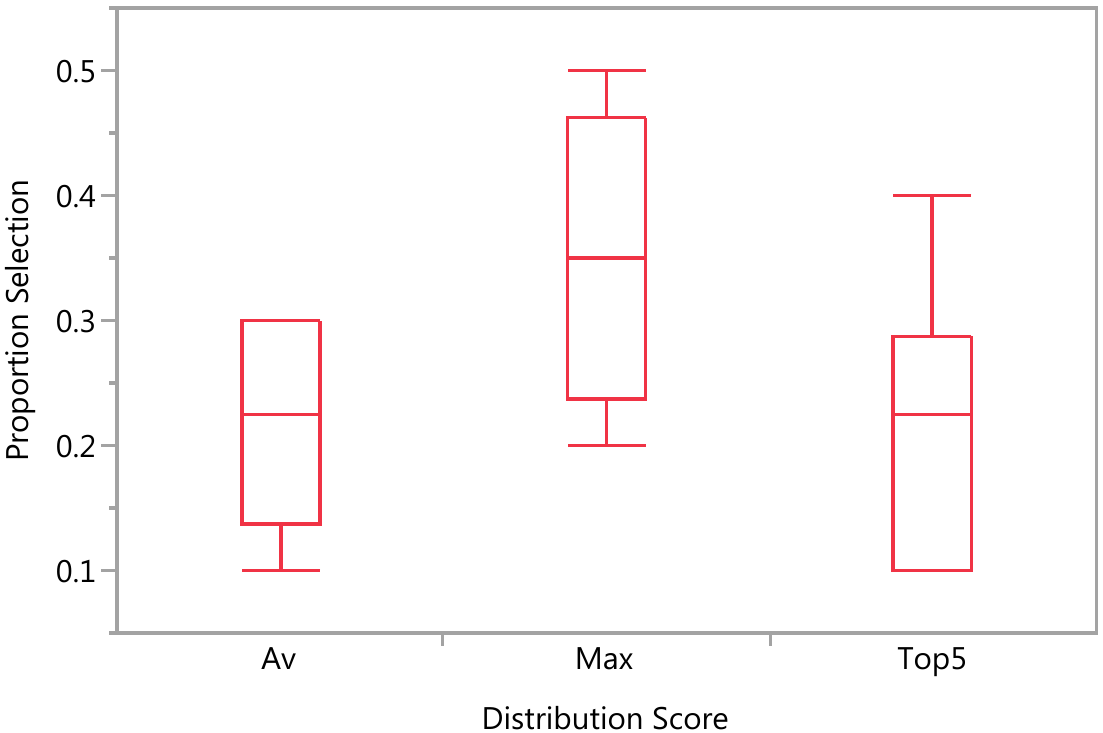}
\caption{Distributions of selection proportions of different R2RS distribution scores across the four projects.}
 \label{fig:RQ2approachh}
\end{figure}

\begin{center}
\fbox{
\begin{minipage}[t]{.95\linewidth}
{\bf RQ2 Summary:}
The selection proportion significantly varies among metrics and metrics families. The TLCC metrics family has the highest average selection proportion, followed by R2RS, R2C, CKJM and SQ.
\end{minipage}
}
\end{center}

\subsection{RQ3: What is the accuracy of single metrics families in predicting impacted classes?}

Table \ref{tab:RQ3newdetailed} reports the average F1 among the 20 datasets achieved by a specific classifier in a specific dataset; the highest accuracy is in bold. Table \ref{tab:Rq3P&R} in the appendix reports results in terms of precision and recall. Table \ref{tab:RQ3newgainF1}  reports the average F1 among different classifiers and projects when using a specific metrics family. We make the following observations:\\
$\bullet$~ R2RS provides an accuracy, in average among classifiers, datasets, and projects, higher than any other single metrics family. The accuracy of other single metrics is between -60\% and -24\% when compared to R2RS.\\
$\bullet$~ In Tika and Isis no combination of classifier and metrics family provides an accuracy higher than 0.2. This suggests that using a single metrics family may provide a very low accuracy. \\
$\bullet$~ R2RS provides the higher accuracy, among other single metrics family, in Ignite and Accumulo. \\

% \begin{table}[!b]
% \includegraphics[width=\columnwidth]{RQ3newdetailed.pdf}
% \caption{Average F1 among the 20 datasets achieved by a specific classifier in a specific dataset. }
% \label{tab:RQ3newdetailed}
% \end{table}

\begin{table*}
\centering

\caption{Average F1 among the 20 datasets achieved by a specific classifier in a specific dataset.}

\label{tab:RQ3newdetailed}
\begin{tabular}{|c|c|c|c|c|c|c|c|c|c|c|}
\hline
\multirow{2}{*}{\begin{tabular}[c]{@{}c@{}}Metric \\ Family\end{tabular}} & \multicolumn{5}{c|}{Accumulo}                           & \multicolumn{5}{c|}{Ignite}                             \\ \cline{2-11}
              & Bagging & J48   & Logistic & Naive Bayes & Rand.Forest & Bagging & J48   & Logistic & Naive Bayes & Rand.Forest \\ \hline
CKJM          & 0.194   & 0.215 & 0.21     & 0.14        & 0.217        & 0.681   & 0.703 & 0.586    & 0.311       & 0.695        \\ \hline
R2C           & 0.198   & 0.189 & 0.172    & 0.19        & 0.196        & 0.189   & 0.225 & 0.169    & 0.223       & 0.177        \\ \hline
R2RS          & \textbf{0.333}   & 0.3   & 0.299    & 0.308       & 0.345        & 0.834   & 0.839 & \textbf{0.906}    & 0.633       & 0.837        \\ \hline
SQ            & 0.216   & 0.214 & 0.235    & 0.078       & 0.204        & 0.643   & 0.617 & 0.511    & 0.281       & 0.662        \\ \hline
TLCC          & 0.304   & 0.317 & 0.133    & 0.105       & 0.296        & 0.563   & 0.573 & 0.527    & 0.748       & 0.575        \\ \hline
\multirow{2}{*}{}              & \multicolumn{5}{c|}{Isis}                               & \multicolumn{5}{c|}{Tika}                               \\ \cline{2-11} 
                               & Bagging & J48   & Logistic & Naive Bayes & Rand.Forest & Bagging & J48   & Logistic & Naive Bayes & Rand.Forest \\ \hline
CKJM          & 0.019   & 0.018 & 0.022    & 0.036       & 0.019        & 0.066   & 0.074 & 0.061    & 0.068       & 0.06         \\ \hline
R2C           & 0.018   & 0.018 & 0.016    & 0.017       & 0.018        & 0.047   & 0.078 & 0.053    & \textbf{0.147}       & 0.045        \\ \hline
R2RS          & 0.019   & 0.018 & 0.017    & 0.017       & 0.019        & 0.049   & 0.037 & 0.041    & 0.019       & 0.039        \\ \hline
SQ            & 0.02    & 0.021 & 0.027    & 0.035       & 0.02         & 0.058   & 0.033 & 0.05     & 0.049       & 0.055        \\ \hline
TLCC          & 0.044   & \textbf{0.045} & 0.03     & 0.043       & 0.039        & 0.062   & 0.011 & 0        & 0.038       & 0.046 \\  \hline     
\end{tabular}
\end{table*}

\begin{table}
\centering
\caption{Average F1 provided by single metrics families among different classifiers, datasets and projects.}
\label{tab:RQ3newgainF1}
\begin{tabular}{|c|c|c|c|}
\hline
                  & \textbf{Average F1} & \textbf{Difference to R2RS} \\ \hline
\textbf{CKJM} & 0.220                    & -26\%                   \\ \hline
\textbf{R2C}   & 0.119                      & -60\%                   \\ \hline
\textbf{R2RS}     & 0.295                    & -                    \\ \hline
\textbf{SQ}     & 0.201                     & -32\%                  \\ \hline
\textbf{TLCC}     & 0.225                     & -24\%                  \\ \hline
\end{tabular}
\end{table}

%Figure \ref{fig:RQ3new} reports distributions of F1, among the five classifiers and the 20 samples (see Section 3.3.2), for each of the four projects, by using a specific metric family.  We make the following observations:\\

%$\bullet$~ In Tika and Isis no combination of classifier and family metrics provides an accuracy higher than 0.2. This suggests that using a single metrics family may provide a very low accuracy. \\
%$\bullet$~ R2RS provides the higher accuracy, among other single metrics family, in Ignite and Accumulo. This suggests the need to use more than one family of metrics to predict with a reasonable accuracy the set of impacted class.\\

%\begin{figure*}[!t]
%   \centering
%    \includegraphics[width=\textwidth,height=.4\paperheight,keepaspectratio]{RQ3new.pdf}
%    \caption{Distributions of F1, achieved by each of the five classifier, across five classifiers and 20 samples of each of the four project, by using specific family metrics (color).}
%    \label{fig:RQ3new}
%\end{figure*}

In order to test H100 we applied the Kruskal-Wallis test comparing, for each of the four projects, the F1 achieved by each classifier using a specific metric family. Our results show a statistical significant difference, i.e., we can reject H10, in all 20 cases (five classifiers over four projects).

\begin{center}
\fbox{
\begin{minipage}[t]{.95\linewidth}
{\bf RQ3 Summary:}
R2RS provides an accuracy, in average among classifiers, datasets, and projects, practically higher than any other metrics family. The accuracy of other single metrics is between -60\% and -24\% when compared to R2RS.
\end{minipage}
}
\end{center}

\goodbreak
\subsection{RQ4: How much does using R2RS improve the accuracy in predicting impacted classes?}

\begin{figure*}[!t]
    \centering
    \includegraphics[width=\textwidth,keepaspectratio]{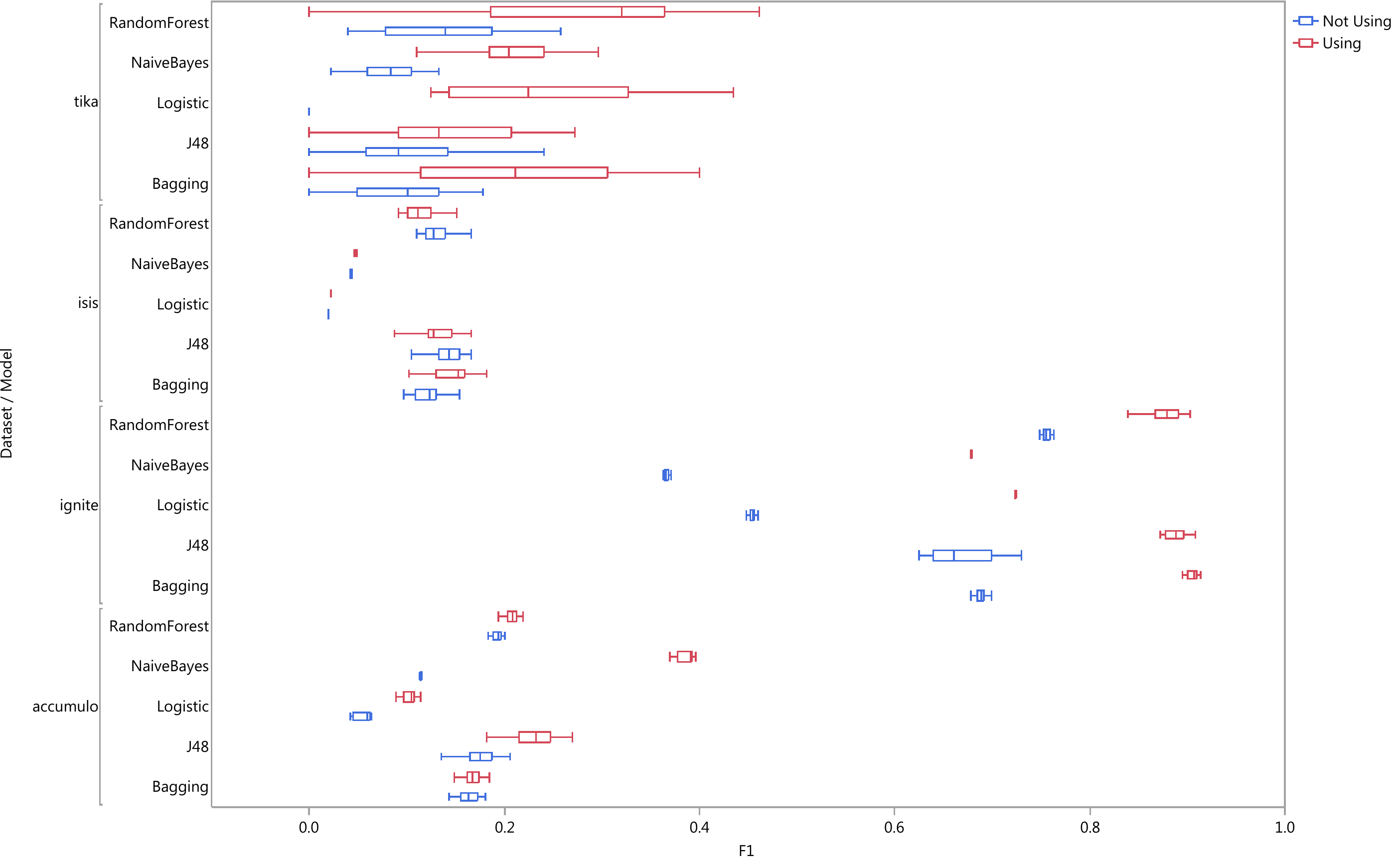}
    \caption{Distributions of F1, achieved by each of the five classifier, across the 20 samples of each of the four project, by using (blue) or not (red) the R2RS metrics.}
    \label{fig:RQ3BWplot}
\end{figure*}

Figure \ref{fig:RQ3BWplot} reports distributions of F1, among the 20 samples (see Section 3.3.2), achieved by each of the five classifiers, for each of the four projects, by using (With) or not using (Without) the R2RS metrics in the classifier. We make the following observations:\\
$\bullet$~ The highest F1 is achieved with R2RS in all 4 projects.\\
$\bullet$~ `With R2RS' provides a higher F1 than `without' in all five classifiers for Tika, Ignite, and Accumulo data sets.\\
$\bullet$~ In ISIS, `with R2RS' provides a lower F1 than `without' in RandomForest and J48 classifiers. However, the F1 of RandomForest and J48 classifiers are lower than the F1 of Bagging `with' R2RS.\\
$\bullet$~ The best classifier varies among projects and according to whether it uses `With R2RS' or `Without'. For instance, in Ignite the best classifier is Bagging `with R2RS' and RandomForest `without R2RS'.\\

Table \ref{tab:RQ3gainF1}  reports the average F1 among different classifiers on the same project when using or not using R2RS metrics; table \ref{tab:Rq4P&R} in the appendix reports results in terms of precision and recall. As we can see the relative gain on F1 ranges from 1\% to 149\%, with an average of 61\% across all projects.

% \begin{figure}[!b]
% \includegraphics[width=\columnwidth]{RQ3gainF1.pdf}
% \caption{Average F1 among different classifiers on the same project when using or not R2RSS metrics.}
%  \label{fig:RQ3gainF1}
% \end{figure}

\begin{table}[b]
\centering
\caption{Average F1 among different classifiers on the same project when using or not R2RSS metrics.}
\label{tab:RQ3gainF1}
\begin{tabular}{|c|c|c|c|}
\hline
                  & \textbf{With} & \textbf{Without} & \textbf{Relative Gain} \\ \hline
\textbf{Accumulo} & 0.219         & 0.141            & 55\%                   \\ \hline
\textbf{Ignite}   & 0.81          & 0.587            & 38\%                   \\ \hline
\textbf{Isis}     & 0.093         & 0.091            & 1\%                    \\ \hline
\textbf{Tika}     & 0.213         & 0.086            & 149\%                  \\ \hline
\end{tabular}
\end{table}

In order to test H110 we applied the Kruskal-Wallis test comparing, for each of the four projects, the F1 achieved by each classifier with and without R2RS metrics. An $x$ in Table \ref{tab:RQ3tests} denotes a p-value $< 0.05$ (i.e., alpha). According to Table \ref{tab:RQ3tests} \noindent \emph{we can reject H110 in 16 out of 20 cases.} The use of R2RS provides a statistically significant improvement in F1 in 80\% of the cases.  The use of R2RS provides a statistically significant improvement in F1 in all cases in Ignite and in three out of four cases in the remaining three projects.

\begin{table}
\centering
\caption{Kruskal-Wallis test comparing, in each of the four projects, the F1 achieved by each classifier with  versus without R2RS metrics. An x denotes a p-value \textless  0.05 (i.e., alpha).}
\label{tab:RQ3tests}
\begin{tabular}{|c|c|c|c|c|c|}
\hline
\textbf{}         & \textbf{Bagging} & \textbf{J48} & \textbf{Logistic} & \textbf{\begin{tabular}[c]{@{}c@{}}Naive\\ Bayes\end{tabular}} & \textbf{\begin{tabular}[c]{@{}c@{}}Random\\ Forest\end{tabular}} \\ \hline
\textbf{Accumulo} &                  & x            & x                 & x                                                              & x                                                                \\ \hline
\textbf{Ignite}   & x                & x            & x                 & x                                                              & x                                                                \\ \hline
\textbf{Isis}     & x                &              & x                 & x                                                              &                                                                  \\ \hline
\textbf{Tika}     & x                &              & x                 & x                                                              & x                                                                \\ \hline
\end{tabular}
\end{table}

\begin{center}
\fbox{
\begin{minipage}[t]{.95\linewidth}
{\bf RQ4 Summary:}
The highest accuracy is achieved with R2RS in all 4 projects. The use of R2RS increases accuracy of an a average of 61\% across projects.
\end{minipage}
}
\end{center}

\newpage

\section{Running Example}
\label{sec:Practice}
\begin{figure}[t]
\includegraphics[width=\columnwidth]{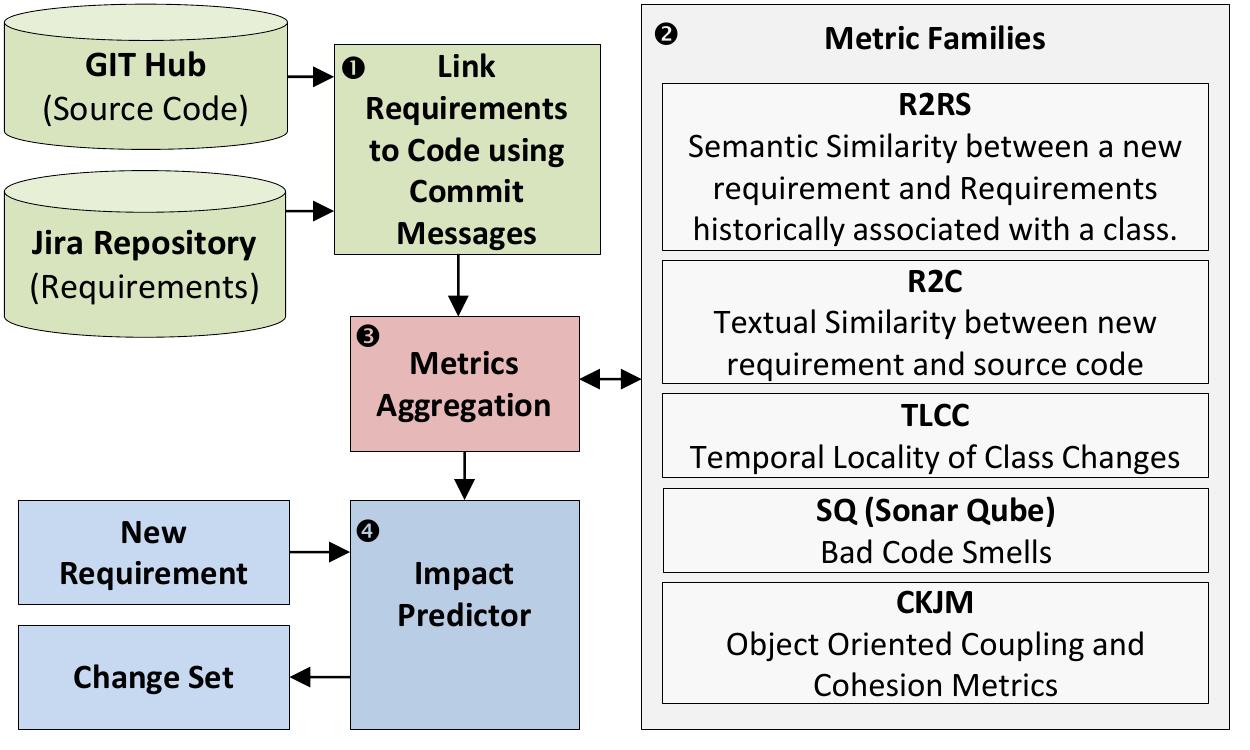}
\caption{An overview of the Impact Prediction Process showing the five metric families from which prediction models are trained.}
\label{fig:Overview}
\end{figure}

In this section, we describe an example of usage of our approach for automated impact prediction. 
Consider Pat as a developer of the Apache Accumulo project (see Section \ref{subsec:exp_obj}). Pat is requested to implement the requirement specified in ACCUMULO-1009\footnote{https://issues.apache.org/jira/browse/ACCUMULO-1009} entitled \emph{Support encryption over the wire} and described as \emph{Need to support encryption between ACCUMULO clients and servers. Also need to encrypt communications between server and servers. Basically need to make it possible for users to enable SSL+thrift.}  In order to implement this requirement, Pat needs to investigate which classes will potentially be modified among all the current 628 classes.
To facilitate this process, Pat uses our automated impact prediction as described in Figure \ref{fig:Overview}. After Pat inputs the new requirement ACCUMULO-1009, the predictor will perform the following steps automatically:
\begin{enumerate}
    \item Collecting historic data from JIRA\footnote{https://issues.apache.org/jira/projects/ACCUMULO/} and Git\footnote{https://github.com/apache/accumulo} ACCUMULO repositories. Specifically, the predictor links previously implemented requirements with code via commit messages.
		\item collecting measures required by the five metrics families.
		\item Aggregating metrics in a format that is readable to WEKA.
		\item Predicting the set of classes impacted by ACCUMULO-1009 using RandomForest trained with the data provided by the previous step.		
\end{enumerate}		

The predictor identifies 41 potentially impacted classes out of the set of 628 classes. Pat analyzes these 41 classes and decides which ones to modify and how. Of these, 17 are actually impacted and 24 are not, leading to recall of approximately 35\% and precision of 41\%. For illustrative purposes we examine the \emph{/core/client/impl} package, which is the package containing the highest number of impacted classes.  The package includes 26 classes of which nine are actually impacted. The predictor correctly identifies four of them (\emph{TabletServerBatchWriter.java}, \emph{ServerClient.java}, \emph{TabletServerBatchReaderIterator.java}, and \emph{Writer.java}) and misses five classes (\emph{MasterClient.java}, \emph{ScannerIterator.java}, \emph{OfflineScanner.java}, \emph{ConditionalWriterImpl.java}, and \emph{ThriftTransportPool.java}).  In addition, the predictor also incorrectly identifies two classes out of six as impacted (\emph{ThriftScanner.java} and \emph{ConnectorImpl.java}).

From a practical perspective this suggests that our automated impact prediction with NLP techniques can certainly not replace the human-intensive task of identifying impacted classes; however, it could potentially serve as a recommendation system that suggests additional classes for the human analyst to inspect. The state of affairs improves if the level of recommendation is raised to the package level.  At the time that ACCUMULO-1009 feature request was issued, Accumulo had 88 packages of which 27 were impacted by the change. If predictions were made for any package containing a potentially impacted class, then the impact predictor would have identified 25 packages of which 16 were actually impacted.  This would have resulted in recall of 59\% and precision of 64\%. It is well known that making predictions at the coarser-grained package level increases accuracy; however, it may be more practical to offer developers a more general, but more accurate, recommendation.

Moreover, we expect that it would be more difficult for a developer to identify impacted classes that are isolated from other impacted classes.  In this example, the \emph{/core/client/impl} package contained nine impacted classes, three packages contained three impacted classes each, and eight packages contained two impacted classes each; however, there were fifteen impacted classes residing on their own in fifteen other distinct packages.  Our approach successfully identified 7 of these isolated classes.  In one additional case, it missed the class, but identified the package, and in the remaining 7 cases we failed to identify either the class or the package. Single class recommendations in this example would have returned precision of 87.5\% (or 100\% at the package level), with recall of 46\% (or 53\% for the package). Recommending any classes or packages that would otherwise have been missed by the developer could be very helpful in practice.

Finally we expect this approach could be particularly useful for developers that are not knowledgeable about the code base because they are recently hired or in cases where the code base is very large and the proportion of impacted class is extremely low (i.e., about 1\%) as were the case in three out of four of our analyzed projects.

%% file: Threats.tex
\section{Threats to Validity}
\label{sec:Threats}
Internal validity refers to whether the experiment is designed with minimal systematic errors so that a causal conclusion drawn from the data is valid. We believe the threats to validity of this type are quite low in this study. In general, software engineering research is an iterative process where validation results support new hypotheses \cite{Basili:1986:ESE:9775.9777}. In this paper we needed to make several decisions such as a specific set of NLP techniques and of approaches to compute a single similarity scores. When making decisions we took into consideration several attributes including past applications and resources. For instance, we choose six textual similarity techniques to measure requirements similarities. These techniques are term matching based comparators considering word semantics using either term distribution, external lexical database WordNet, POS tags, or term sequences. For practical purpose, we avoid heavy weight NLP techniques that would take considerably more computation resources. We note that such decisions are not exactly threats to validity because they can bias the R2RS performance only in a  negative way.

Construct validity generally concerns the degree that the measurements correspond to the construct of interest. Our claim in this paper is that we predict the impact of requirements upon source code.  The datasets we selected utilize JIRA repositories to collect issues. Because bug-fixes are very different in nature to requirements and a classifier designed to predict their impact might be quite different to one designed to predict requirements impact, we selected only issues which were manually tagged by users as feature requests. Our requirements, being from open source projects, differ from industrial projects in the way they are represented, i.e., as informal requests versus traditional user stories or ''shall" statements, and also in the way they relate to each other. Specifically, industrial requirements are typically subject to analysis for avoiding duplicates or inconsistencies \cite{Falessi2017, falessi2013empirical}; we defer to future studies a validation to industrial requirements. We followed a systematic explicit data mining process for determining whether a class had been impacted by a requirement.  Moreover, we selected datasets by maximizing their linkage, diversity, and size (see Table \ref{tab:projects}). Such datasets contains high and varied data points which prevent over-training of the impact predictor, thus improving the reliability of our approach. On the other hand, our requirements are taken from open source projects, and differ in nature from the types of requirements often written for closed-source projects. To provide insights into this, we included sample requirements for each of our projects.
A threat to construct validity related to RQ3 could be that the number of samples (i.e., 20) is low and hence results biased by random effects included while sampling. To check the absence of random effects due to sampling we computed the standard deviation achieved by the same prediction model (i.e., same feature and classifier) on the same dataset over the 20 samples. The standard deviation on F1 resulted in an average of 0.027 and therefore we can assert that very low random effects have impacted our results.
Finally, our approach has been inclusive and hence to consider a large number of complexity metrics such as WMC, even if without a strong evidence of a relation to change proneness. In order to avoid that the use of potentially weak metrics could have favored R2RS, in all three research questions we have adopted an experimental design where R2RS metrics results cannot be impacted by weak metrics. Specifically, the fact that size resulted with a lower IGR and a lower selection frequency than 11 out of 18 R2RS metrics cannot be caused by the presence of a possible weak metric. Similarly, the accuracy gain caused by using R2RS metrics observed in RQ3 cannot be caused by the presence of a possible weak metric in other metric families.

Conclusion validity refers to the degree to which the conclusions are reasonable and reliable. Regarding RQ2, the output of the selection strategy could differ based on the algorithm or the parameters used. We mitigated this threat by using default algorithms and parameters.
Regarding the validation strategy in RQ3, rather than a standard ten-fold validation method, we partitioned the datasets into two fixed sets of 80\% for training and 20\% for test while preserving the order of the timeline. This is due to the temporal nature of two of the feature families, i.e., R2RS and TLCC. 

External validity refers to the generalizability of the approach.  The main threat to validity is that results are related to four projects, all of which are open-source. For selecting projects we used the guidelines provided by Nagappan et al. \cite{nagappan2013diversity}. 
According to Table \ref{tab:projects}, project Tika is small while project Ignite has a very high proportion (57\%) of impacted classes. Thus it could be questionable if these two systems are representative projects. However, we note that no system can be claimed more or less representative than another if we do not know the definitive characteristics of the population we want to generalize results to. In fact, the choice of these two systems matches the suggestion provided by Nagappan et al. \cite{nagappan2013diversity} to include high diversity in the chosen projects. Specifically, the use of Tika and Ignite extends our results to include a discussion of small projects and projects with a high proportion of impacted classes, respectively.
Moreover, in order to support replicability, we provide our datasets, arff files, and weka output as explained in Section \ref{sec:ExperimentalDesign}. Finally, to further support repeatability of our approach, Table \ref{tab:WekaParams} reports the parameters used to train each of the classifiers.  
 

%% file: Related.tex
\section{Related Work}
\label{sec:Related}

\subsection{Change Impact Analysis}
Change impact analysis was defined by Horowitz et al. in 1986 as ``examination of an impact to determine its parts or elements'' \cite{horqwitz1986sodos}. Steffen Lehnert \cite{Lehnert2011} presents a literature review of 150 studies and a taxonomy of change impact analysis. His study highlights the many different ways in which change impact analysis can be used and the diverse stages of a project at which it can be applied. For instance, during early-stage activities, change impact analysis can help identify impacts of requirements changes on other requirements \cite{1544761, GOKNIL2014950}. During late-stage activities, change impact analysis can support the quality and correctness of software architectures \cite{williams2010characterizing}. Change impact analysis is also beneficial when an engineer is provided with high-level change requests from a customer,  and has to decide whether or not a change is feasible, and hence has to estimate the cost and risk of a change \cite{O'Neal:2003, walker2006lightweight}. 
Li et al. \cite{li2013survey} present a survey of code-based change impact analysis techniques. As explained by Li et al., change impact analysis aims at identifying the ripple effects and preventing side effects, i.e., the likelihood that a change to a particular module may cause problems in the rest of the program \cite{rovegaard2008empirical}. Thus, our aim to predict the set of classes impacted by a requirement  fully fits neither the concept of change impact analysis used in the current literature nor the framework for change impact analysis study classification proposed by Li et al. \cite{li2013survey}. Several change impact analysis studies focus on code properties \cite{li2013survey} such as static or dynamic dependence analysis and in particular on co-changes \cite{rolfsnes2016generalizing, d2009visualizing, kagdi2013integrating, robillard2008topology, poshyvanyk2007feature}. 
In 2004, Ying et al. used association rule mining to identify which classes were frequently modified together in previous releases \cite{ying2004predicting}. Using a Frequency Pattern Tree (FPTree) and a low minimum-support threshold, they developed an association rule algorithm and ran it on the Eclipse and Mozilla projects. The developer identified an initial file to be changed, and the association rules were used to recommend other impacted classes.  In 2005, Zimmermann et al. also used association rule mining techniques to mine version archives for change recommendation purposes \cite{ZimmermannTSE2005}. Instead of recommending files, their approach suggests fine-grained program entities such as functions or variables to be changed. One benefit of mining change associations from archives is that the change prediction is not reliant on a specific programming language; however, as Zimmermann et al. pointed out \cite{ZimmermannTSE2005}, the quality of the prediction is directly dependent on the developer's past practices. In 2011, Eski et al. \cite{eski2011empirical} proposed a history based approach in which different types of code modifications, such as a new field or a modified method, were weighted differently.  Classes with higher historical weightings were predicted to be more likely to be changed in the future.  Association rule mining could be integrated with our approach to enable the identification of additional, currently missed, classes.
Jiang et al. \cite{jiang2017programmers} recently provide evidence about the non use of any change impact analysis tool support  by practitioners. This suggests that further effort should be spent on this area.

\subsubsection{Feature Location}
Feature location is a particular type of impact analysis which aims at identifying portions of code that implemented a requirement. Note that in our paper we are interested in identifying portions of code that will implement a requirement. 
Dit et al. \cite{dit2013feature} provide a taxonomy and a survey about studies proposing feature location techniques. 
Dit et al. \cite{dit2013feature} characterize previous studies according to the combination of static, dynamic, and textual techniques used. In this paper we use a combination of static (CKJM, SQ, TLCC) and textual (R2C, R2RS) techniques. 
It is worth noting that both Dit et al. \cite{dit2013feature} and Li et al. \cite{li2013survey} agree on the lack of benchmarks. The present study tries to cover this lack by using a semi-automated and fully replicable measurement and analysis procedure. 
Eisenbarth et al. \cite{eisenbarth2003locating} presented a technique combining dynamic and static analyses to rapidly focus on the system parts that relate to a specific set of features. Their technique is particularly suitable when the number of features to locate is higher than one.
Antoniol et al. \cite{DBLP:journals/tse/AntoniolCCLM02} used a probabilistic approach to retrieve trace links between code and documentation. Hayes et al. used the Vector Space Model (VSM) algorithm in conjunction with a thesaurus to establish trace links \cite{DBLP:journals/tse/HayesDS06}.  In general, these studies observed high Recall and low Precision, which led to subsequent research effort focusing on  precision. For example, in order to understand the semantics or context of elements in the artifacts some studies applied  Latent Semantic Indexing \cite{DeLucia:ArtefManag} and Latent Dirichlet Allocation (LDA) \cite{DBLP:conf/re/DekhtyarHSHD07, DBLP:conf/icse/AsuncionAT10}. Moreover, researchers combined the results of individual algorithms \cite{lohar2013improving, DBLP:conf/re/DekhtyarHSHD07, GethersICSM}, applied AI swarm techniques \cite{DBLP:journals/re/SultanovHK11}, and combined heuristic rules with trace retrieval techniques \cite{DBLP:journals/jss/SpanoudakisZPK04, 6636704}. Our approach differs from trace link generation techniques because it leverages other information such as historical changes and therefore goes far beyond textual or semantic similarity in identifying a set of impacted classes.
More recently, in 2015, Acharya et al. performed a study on static program slicing and its usefulness in impact analysis \cite{acharya2011practical}. Static program slicing is the process of isolating or ``slicing'' the code in order to extract parts of the code related to specific functions. Their tool is tightly integrated to the CodeSurfer API and Visual Studio and is therefore not suited for more general use.

\subsubsection{Bug Reports to Code}

In 2005, Canfora et. al used information retrieval techniques to identify files that were created or changed, as a result of the creation of Bugzilla tickets \cite{canfora2005impact}. For a new change request, they identified files impacted by similar past change requests. They used standard NLP techniques such as stemming words, computing word frequencies, and scoring the documents by relevance using the new bugzilla ticket as a search query.

In 2013, Kim et al. worked on a system to predict what source code files would change as a result of bug-fix requests \cite{kim2013should}. They used the Mozilla FireFox and Core code repositories as their corpus in tandem with the public Bugzilla database for both. They built what was essentially a two-pass classifier. For the first pass, they hand-tagged bug-fix requests as either ``USABLE'' or ``NOT USABLE'', and then trained a classifier based on natural language processing metrics to predict whether the bug-fix request was usable. On most of the analyzed datasets at least half of the bug-fixes were filtered out as overly vague. In the second pass, they extracted and stemmed words from each bug-fix request, computed term frequencies, and filtered out all common stop-words. They then used all of this information to create metrics for classification, along with the metadata of the bug-fix request (system, epic, etc.), and then trained the classifier. They claimed 70\% accuracy on data that they defined as usable \cite{kim2013should}.

In particular, our research most closely relates to the studies of Canfora et al.\cite{canfora2005impact} and Kim et al.,\cite{kim2013should} in that we use change request history and NLP similarity measures to identify a set of impacted classes. Like Tsantalis et al. \cite{tsantalis2005predicting}, we also use class dependency and quality analysis in the prediction process.  However, our work focuses on new features instead of bug fixes and proposes and evaluates multiple techniques for computing similarity between requirements and code.  We also integrate this approach with other families of techniques.

\subsection{Change Proneness}
Several studies investigated the probability that a class would change in the future. Our work differs form these studies because they do not take into account the number and type of requirements that still have to be implemented.
Bieman et al. \cite{bieman2003design} investigated the relation between change proneness and design patterns. They also observed that size impacts change proneness.
Khomh et al. \cite{khomh2009exploratory} show a positive correlation between smells and change proneness.
Zhou et al. \cite{zhou2009examining} analyzed the potentially confounding effect of size on the correlation between complexity and change proneness. Their results show that complexity seems correlated with change proneness mainly because complexity is correlated which size which is in turn correlated with change proneness. Thus, they advise to take into account size every time the correlation between complexity and change proneness is investigated.
Arvanitou et al. \cite{arvanitou2017method} very recently proposed a new approach to predict the change proneness of a class that takes into account temporal locality and complexity of the current and the connected classes. Differently from our approach, they do not consider any information related to the similarity to the code or the similarity with future or past implemented requirements.

\subsection{Impacted Class Prediction}
Lindvall and Sandahl performed a study on manual, requirements driven impact analysis inspired by Bohner's previous work \cite{lindvall1998well}. They worked with developers of the PMR project over two releases and four years, recording relevant data for impact analysis. Based on the given requirements for a release, the developers predicted which C++ classes in the code-base were likely to change. They observed an average Recall of 0.76 and a Precision of 0.96 %(i.e., F2 of 0.79). 
However we note that their projects were order of magnitude smaller and extremely more balanced than our projects. Therefore, it is impossible to argue whether our approach would have performed better than the humans in the Lindvall and Sandahl study. Furthermore, automation offers potential value for larger sized projects. 
Tsantalis et al. developed a tool to predict the probability of change for a class using a system's code base and its change history. Given a class's dependencies, summarized by its CKJM metric scores and its history of changes, their method used logistical regression to predict the likelihood of a change \cite{tsantalis2005predicting}.
 

%% file: Conclusion.tex
\section{Conclusion}
\label{sec:Conclusion}
In this paper we have presented a new family of features, named Requirements to Requirements Set (R2RS), which is designed to measure the semantic similarity between a new requirement and the set of requirements previously associated with a source code class. We comparatively evaluated 18 R2RS metrics against and in combination with, 16 other metrics belonging to four other types of information such as temporal locality, similarity to code, CKJM, and code smells.  Our experimental evaluation was conducted using five classifiers and 78 releases belonging to four large open-source projects.
Performing an exhaustive search to understand if R2RS metrics are useful is unfeasible. Therefore, we proceeded by applying three complementary methods which are based on: 1) Information Gain Ratio, 2) selection for maximize accuracy, and 3) gain in accuracy when classifiers use R2RS metrics. A novelty of this paper is the use of IGR and selection frequency for validating the prediction power of a new metric. IGR and selection frequency have, to best of our knowledge, never been applied in software engineering studies, and they stem from solid machine learning studies \cite{Gray:1990, Mitchell:1997, harris2002information,john1994irrelevant, kohavi1997wrappers}.
To summarize, experimental results suggest that: 
\begin{enumerate}
    \item TLCC\_Lin, i.e., a simple class change frequency, provided the highest IGR, and was selected a higher proportion of times when compared to the remaining 33 metrics. Given this result and the the simplicity and easiness of computing TLCC\_Lin, we suggest starting with this metric when in need to predict which class will change.
    \item The accuracy, i.e., F1, in average among classifiers and projects, of other single metrics is between -60\% and -24\% when compared to R2RS.  
    \item Using R2RS metrics increases prediction accuracy, from 1\% to 149\%, with an a average of 61\%, among projects. Moreover, the highest accuracy is achieved by leveraging R2RS in all four projects.
    \item Eleven out of eighteen R2RS metrics have both a higher IGR and a higher selection proportion than well established change proneness metrics such as size (SQ\_NCLOC) and smells (SQ\_Viol) \cite{khomh2009exploratory, arvanitou2017method, zhou2009examining}. This means that focused information, about the specific requirement to implement in the future, outperformed coarse grained information such as size and smells. Thus, the likelihood of a class to change depends more on the content of the requirement than on class attributes such as size. This result is in contrast to a recent previous study showing that "size is the only unique metric" for software prediction \cite{Gil:2017:CSM:3135854.3135904}. 
	\item Distribution scores, among the last ten requirements touching the class, and NLP techniques of R2RS metrics slightly differ in IGR and selection proportion. The Max distribution score returned highest IGR and  selection proportion when compared to the average or the top five similarity scores.
	
\end{enumerate}
In conclusion, our results show that leveraging semantic similarities between new and previously implemented requirements generally resulted in more accurate predictions. However, future research effort is needed in the following areas:
\begin{itemize}
\item \textit{Identifying dataset characteristics favoring R2RS}. This paper is based on only four projects and therefore, as with the case of any empirical work, it does not aim to provide completely generalizable results. However, some results were consistent across projects and therefore likely to be generalizable. If we could analyze hundreds of thousands of projects we could discover project characteristics that influence the accuracy of R2RS and predict in advance the best combination of features to use \cite{DBLP:conf/msr/0004RCRHV16}. Mining data from each individual project for analysis is extremely time consuming, which limited our current study to four projects. 
%In the current work, we However, analyzing four projects is already challenging in terms of dataset collection (i.e., several hundreds of millions of predictions of requirement-class changes stored in more than 10GB of data. Therefore, this could be not practically possible.
\item \textit{Understanding when class change prediction is feasible}. There is a significant difference between the accuracy and usefulness of predictions across projects \cite{falessi2013empirical}. Our four projects differ in several characteristics including percentage of touched classes. Projects with a very low percentage of positives and a very high number of classes are intuitively hard to predict for our approach, however this task is even harder for humans to perform manually. User studies are therefore needed to provide insights into when and where automated techniques are useful to humans.  We expect to find a break even point where automation becomes more accurate than fully manual techniques. 
\item \textit{Understanding the utility of class change prediction}. The ability to predict which classes are impacted by a new requirement can potentially support a slew of tasks including refactoring decisions, defect prediction, and effort estimation. An increased understanding of such tasks could enable us to build more effective, contextualized prediction algorithms. 
\item \textit{Enhancing and modifying feature families}. Our study encompasses 34 different features, which is a large number when compared to the state of the art. However, there are features such as association rules or software architecture metrics that were not considered and could potentially improve accuracy.  Furthermore, there is value in understanding whether any features within each feature family are detrimental to prediction accuracy and should be omitted.  
\item \textit{Leveraging R2RS accuracy as a proxy for software architecture and requirements quality}. In closing, we note that a scenario in which TCLL is more valuable than R2RS likely imply that some (non low-level service) classes are touched regardless of the semantic of the change request; this could be an indicator of poor architectural design. Software applications are designed to promote several quality attributes, such as  maintainability and extensibility, through the separation of concerns \cite{Falessi:2011:DTS:1978802.1978812}. A project in which R2RS returns accurate results could indicate a clear separation of concerns at both the architecture and requirements level. This reasoning is as yet unproven but represents a novel opportunity to explore synergies between requirements specification, architectural design, impact prediction, and natural language processing.
\end{itemize}

%% file: Appendix.tex
\newpage
\section{Appendix}

\begin{table*}[!t]
    \centering
    \small\addtolength{\tabcolsep}{-4pt}
    \begin{tabular}{|l|c|c|c|c|c|c|c|c|c|c|c|c|c|c|c|c|c|c|c|c|}
\hline
&\multicolumn{5}{c|}{\bf Accumulo}&\multicolumn{5}{c|}{\bf Ignite}&\multicolumn{5}{c|}{\bf Isis}&\multicolumn{5}{c|}{\bf Tika}\\ \cline{2-21}
&Bag&J48&Log&NB&RF&Bag&J48&Log&NB&RF&Bag&J48&Log&NB&RF&Bag&J48&Log&NB&RF\\ \hline
$CKJM\_CA$& &$\bullet$& & & & & &$\bullet$& & & &$\bullet$& & & & & & & & \\ \hline
$CKJM\_CBO$& & & & & & & &$\bullet$& & & &$\bullet$& & & & & & & & \\ \hline
$CKJM\_DIT$&$\bullet$& &$\bullet$& &$\bullet$& & &$\bullet$&$\bullet$& &$\bullet$&$\bullet$& & &$\bullet$& & & & & \\ \hline
$CKJM\_LCOM$& & & & & & & & & & &$\bullet$& & & &$\bullet$& & & & & \\ \hline
$CKJM\_NOC$& & & &$\bullet$& &$\bullet$& & & & & &$\bullet$& & & & & & & & \\ \hline
$CKJM\_NPM$& & & & & & & &$\bullet$& & &$\bullet$&$\bullet$& & &$\bullet$& & & & & \\ \hline
$CKJM\_RFC$& & & & & & & &$\bullet$& & &$\bullet$&$\bullet$& & &$\bullet$& & & & & \\ \hline
$CKJM\_WMC$& & & & & & & &$\bullet$& & & & & & & & & & & & \\ \hline
$R2C\_JSD$&$\bullet$& &$\bullet$& & &$\bullet$& & & &$\bullet$&$\bullet$& & & &$\bullet$& & & & & \\ \hline
$R2C\_VSM$& & & & & & & &$\bullet$& & & &$\bullet$& & & & & & & & \\ \hline
$R2RS\_CMC\_Av$& & &$\bullet$& & & &$\bullet$&$\bullet$&$\bullet$& & &$\bullet$& & & & & & & & \\ \hline
$R2RS\_CMC\_Max$&$\bullet$&$\bullet$& &$\bullet$&$\bullet$&$\bullet$&$\bullet$&$\bullet$&$\bullet$&$\bullet$& & & & & & & & & & \\ \hline
$R2RS\_CMC\_Top5$& &$\bullet$&$\bullet$& &$\bullet$&$\bullet$& &$\bullet$& &$\bullet$&$\bullet$& & & &$\bullet$& & & & & \\ \hline
$R2RS\_JSD\_Av$& & &$\bullet$& & & & & & &$\bullet$&$\bullet$& &$\bullet$&$\bullet$&$\bullet$& & & & & \\ \hline
$R2RS\_JSD\_Max$&$\bullet$&$\bullet$& & &$\bullet$&$\bullet$&$\bullet$& & &$\bullet$& & & &$\bullet$& & & & & & \\ \hline
$R2RS\_JSD\_Top5$& &$\bullet$& & & &$\bullet$&$\bullet$& & &$\bullet$& & & &$\bullet$& & & & & & \\ \hline
$R2RS\_BC\_Av$& & & & &$\bullet$& & &$\bullet$& & & & & & & & & & & & \\ \hline
$R2RS\_BC\_Max$&$\bullet$& & & &$\bullet$& & & & & &$\bullet$& &$\bullet$& &$\bullet$& & & & & \\ \hline
$R2RS\_BC\_Top5$& &$\bullet$& &$\bullet$&$\bullet$& & &$\bullet$& &$\bullet$& & & & & & & & & & \\ \hline
$R2RS\_GC\_Av$&$\bullet$& & & & & & & & &$\bullet$&$\bullet$& & & &$\bullet$& & & & & \\ \hline
$R2RS\_GC\_Max$&$\bullet$& &$\bullet$& &$\bullet$& & & & & & &$\bullet$& & & & & & & & \\ \hline
$R2RS\_GC\_Top5$& &$\bullet$& & & & & & & & &$\bullet$& & & & & & & & & \\ \hline
$R2RS\_VSM\_Av$&$\bullet$& &$\bullet$& &$\bullet$& & &$\bullet$& & & &$\bullet$& &$\bullet$& & & & & & \\ \hline
$R2RS\_VSM\_Max$&$\bullet$&$\bullet$& & &$\bullet$&$\bullet$&$\bullet$& & & &$\bullet$& & & &$\bullet$& & & & & \\ \hline
$R2RS\_VSM\_Top5$& &$\bullet$& &$\bullet$& &$\bullet$& &$\bullet$& & & & & & & & & & & & \\ \hline
$R2RS\_OPC\_Av$& & &$\bullet$&$\bullet$& & &$\bullet$& & & & & & & & & & & & & \\ \hline
$R2RS\_OPC\_Max$&$\bullet$&$\bullet$& &$\bullet$&$\bullet$&$\bullet$& &$\bullet$& &$\bullet$&$\bullet$&$\bullet$& & &$\bullet$& &$\bullet$& & & \\ \hline
$R2RS\_OPC\_Top5$& & & & & & & & & &$\bullet$& & & &$\bullet$& & & & & & \\ \hline
$SQ\_Com$& & & & & & & &$\bullet$& & & &$\bullet$& & & & & & & & \\ \hline
$SQ\_NCLOC$& & & & & & & &$\bullet$& & &$\bullet$&$\bullet$& & &$\bullet$& & & & & \\ \hline
$SQ\_Viol$& &$\bullet$& & & & & &$\bullet$& & & &$\bullet$& & & & & & & & \\ \hline
$TLCC\_Lin$&$\bullet$& & & &$\bullet$& & & & &$\bullet$&$\bullet$&$\bullet$& &$\bullet$&$\bullet$& &$\bullet$&$\bullet$&$\bullet$&$\bullet$\\ \hline
$TLCC\_Log$& & & &$\bullet$& &$\bullet$&$\bullet$&$\bullet$& & & &$\bullet$&$\bullet$& & & & & & &$\bullet$\\ \hline
$TLCC\_SCP$&$\bullet$&$\bullet$& & &$\bullet$&$\bullet$&$\bullet$& & &$\bullet$&$\bullet$&$\bullet$& & & &$\bullet$&$\bullet$& & &$\bullet$\\ \hline
   \end{tabular}
    \caption{Metrics selected per classifier for each project.}
    \label{tab:MetricsSelectionLarge}
\end{table*}

%\begin{figure*}[!b]
%\includegraphics[width=0.9\textwidth]{data.pdf}
%\caption{Which metric has been selected by which classifier in which project.}
% \label{fig:MetricsSelection}
%\end{figure*}

%\begin{figure*}[!b]
%\includegraphics[width=\textwidth]{data2.pdf}
%\caption{Which metric has been selected by which classifier in which project.}
% \label{fig:MetricsSelection}
%\end{figure*}

\begin{table*}[!b]
    \centering
    \tiny\addtolength{\tabcolsep}{-4pt}
    \begin{tabular}{|l|l|l|l|l|l|l|l|l|l|}
\hline
\multicolumn{5}{|c|}{\bf \small{Accumulo}}&\multicolumn{5}{c|}{\bf \small{Ignite}}\\ \hline
Bag&J48&Log&NB&RF&Bag&J48&Log&NB&RF\\ \hline
CKJM\_DIT&CKJM\_CA&CKJM\_DIT&CKJM\_NOC&CKJM\_DIT&CKJM\_NOC&R2RS\_JSD\_Max&CKJM\_CA&CKJM\_DIT&R2C\_JSD\\ \hline
R2C\_JSD&R2RS\_CMC\_Max&R2C\_JSD&R2RS\_CMC\_Max&R2RS\_CMC\_Max&R2C\_JSD&R2RS\_CMC\_Max&CKJM\_CBO&R2RS\_CMC\_Av&R2RS\_CMC\_Max\\ \hline
R2RS\_CMC\_Max&R2RS\_CMC\_Top5&R2RS\_CMC\_Av&R2RS\_BC\_Top5&R2RS\_CMC\_Top5&R2RS\_CMC\_Max&R2RS\_JSD\_Top5&CKJM\_DIT&R2RS\_CMC\_Max&R2RS\_CMC\_Top5\\ \hline
R2RS\_JSD\_Max&R2RS\_JSD\_Max&R2RS\_CMC\_Top5&R2RS\_VSM\_Top5&R2RS\_JSD\_Max&R2RS\_CMC\_Top5&R2RS\_CMC\_Av&CKJM\_NPM&&R2RS\_JSD\_Av\\ \hline
R2RS\_BC\_Max&R2RS\_JSD\_Top5&R2RS\_JSD\_Av&R2RS\_OPC\_Av&R2RS\_BC\_Av&R2RS\_JSD\_Max&R2RS\_VSM\_Max&CKJM\_RFC&&R2RS\_JSD\_Max\\ \hline
R2RS\_GC\_Av&R2RS\_BC\_Top5&R2RS\_GC\_Max&R2RS\_OPC\_Max&R2RS\_BC\_Max&R2RS\_JSD\_Top5&R2RS\_OPC\_Av&CKJM\_WMC&&R2RS\_JSD\_Top5\\ \hline
R2RS\_GC\_Max&R2RS\_GC\_Top5&R2RS\_VSM\_Av&TLCC\_Log&R2RS\_BC\_Top5&R2RS\_VSM\_Max&TLCC\_Log&R2C\_VSM&&R2RS\_BC\_Top5\\ \hline
R2RS\_VSM\_Av&R2RS\_VSM\_Max&R2RS\_OPC\_Av&&R2RS\_GC\_Max&R2RS\_VSM\_Top5&TLCC\_SCP&R2RS\_CMC\_Av&&R2RS\_GC\_Av\\ \hline
R2RS\_VSM\_Max&R2RS\_VSM\_Top5&&&R2RS\_VSM\_Av&R2RS\_OPC\_Max&&R2RS\_CMC\_Max&&R2RS\_OPC\_Max\\ \hline
R2RS\_OPC\_Max&R2RS\_OPC\_Max&&&R2RS\_VSM\_Max&TLCC\_Log&&R2RS\_CMC\_Top5&&R2RS\_OPC\_Top5\\ \hline
TLCC\_Lin&SQ\_Viol&&&R2RS\_OPC\_Max&TLCC\_SCP&&R2RS\_BC\_Av&&TLCC\_Lin\\ \hline
TLCC\_SCP&TLCC\_SCP&&&TLCC\_Lin&&&R2RS\_BC\_Top5&&TLCC\_SCP\\ \hline
&&&&TLCC\_SCP&&&R2RS\_VSM\_Av&&\\ \hline
&&&&&&&R2RS\_VSM\_Top5&&\\ \hline
&&&&&&&R2RS\_OPC\_Max&&\\ \hline
&&&&&&&SQ\_Com&&\\ \hline
&&&&&&&SQ\_NCLOC&&\\ \hline
&&&&&&&SQ\_Viol&&\\ \hline
&&&&&&&TLCC\_Log&&\\ \hline
\multicolumn{10}{c}{}\\ \hline
\multicolumn{5}{|c|}{\bf \small{Isis}}&\multicolumn{5}{c|}{\bf \small{Tika}}\\ \hline
Bag&J48&Log&NB&RF&Bag&J48&Log&NB&RF\\ \hline
Bag&J48&Log&NB&RF&Bag&J48&Log&NB&RF\\ \hline
R2C\_JSD&CKJM\_CA&R2RS\_JSD\_Av&R2RS\_JSD\_Av&CKJM\_DIT&TLCC\_SCP&R2RS\_OPC\_Max&TLCC\_Lin&TLCC\_Lin&TLCC\_Lin\\ \hline
R2RS\_BC\_Max&CKJM\_CBO&R2RS\_BC\_Max&R2RS\_JSD\_Max&CKJM\_LCOM&&TLCC\_Lin&&&TLCC\_Log\\ \hline
R2RS\_CMC\_Top5&CKJM\_DIT&TLCC\_Log&R2RS\_JSD\_Top5&CKJM\_NPM&&TLCC\_SCP&&&TLCC\_SCP\\ \hline
R2RS\_GC\_Av&CKJM\_NOC&&R2RS\_VSM\_Av&CKJM\_RFC&&&&&\\ \hline
R2RS\_GC\_Top5&CKJM\_NPM&&R2RS\_OPC\_Top5&R2C\_JSD&&&&&\\ \hline
R2RS\_OPC\_Max&CKJM\_RFC&&TLCC\_Lin&R2RS\_CMC\_Top5&&&&&\\ \hline
R2RS\_JSD\_Av&R2C\_VSM&&&R2RS\_JSD\_Av&&&&&\\ \hline
R2RS\_VSM\_Max&R2RS\_CMC\_Av&&&R2RS\_BC\_Max&&&&&\\ \hline
TLCC\_SCP&R2RS\_GC\_Max&&&R2RS\_GC\_Av&&&&&\\ \hline
TLCC\_Lin&R2RS\_VSM\_Av&&&R2RS\_VSM\_Max&&&&&\\ \hline
SQ\_NCLOC&R2RS\_OPC\_Max&&&R2RS\_OPC\_Max&&&&&\\ \hline
CKJM\_RFC&SQ\_Com&&&SQ\_NCLOC&&&&&\\ \hline
CKJM\_DIT&SQ\_NCLOC&&&TLCC\_Lin&&&&&\\ \hline
CKJM\_LCOM&SQ\_Viol&&&&&&&&\\ \hline
CKJM\_NPM&TLCC\_Lin&&&&&&&&\\ \hline
&TLCC\_Log&&&&&&&&\\ \hline
&TLCC\_SCP&&&&&&&&\\ \hline

   \end{tabular}
    \caption{Metrics selected per classifier for each project.}
    \label{tab:MetricsSelectionCompact}
\end{table*}

\begin{table*}[]
\centering

\label{tab:pre_rec_rq3}
\begin{tabular}{|l|l|l|l|l||l|l|l|l|l|}
\hline
\textbf{Dataset}  & \textbf{Classifier} & \textbf{MetricFamily} & \textbf{Precision} & \textbf{Recall} & \textbf{Dataset} & \textbf{Classifier} & \textbf{MetricFamily} & \textbf{Precision} & \textbf{Recall} \\ \hline

\textbf{accumulo}&Bagging&CKJM&0.124&0.437&\textbf{isis}&Bagging&CKJM&0.01&0.557\\ \hline
\textbf{accumulo}&Bagging&R2C&0.119&0.59&\textbf{isis}&Bagging&R2C&0.009&0.551\\ \hline
\textbf{accumulo}&Bagging&R2RS&0.23&0.606&\textbf{isis}&Bagging&R2RS&0.01&0.632\\ \hline
\textbf{accumulo}&Bagging&SQ&0.138&0.496&\textbf{isis}&Bagging&SQ&0.01&0.557\\ \hline
\textbf{accumulo}&Bagging&TLCC&0.213&0.531&\textbf{isis}&Bagging&TLCC&0.023&0.398\\ \hline
\textbf{accumulo}&BayesNet&CKJM&0.12&0.323&\textbf{isis}&BayesNet&CKJM&0.013&0.468\\ \hline
\textbf{accumulo}&BayesNet&R2C&0.112&0.537&\textbf{isis}&BayesNet&R2C&0.01&0.539\\ \hline
\textbf{accumulo}&BayesNet&R2RS&0.299&0.573&\textbf{isis}&BayesNet&R2RS&0.009&0.592\\ \hline
\textbf{accumulo}&BayesNet&SQ&0.141&0.408&\textbf{isis}&BayesNet&SQ&0.013&0.525\\ \hline
\textbf{accumulo}&BayesNet&TLCC&0.248&0.536&\textbf{isis}&BayesNet&TLCC&0.025&0.388\\ \hline
\textbf{accumulo}&J48&CKJM&0.144&0.447&\textbf{isis}&J48&CKJM&0.009&0.543\\ \hline
\textbf{accumulo}&J48&R2C&0.113&0.575&\textbf{isis}&J48&R2C&0.009&0.573\\ \hline
\textbf{accumulo}&J48&R2RS&0.197&0.631&\textbf{isis}&J48&R2RS&0.009&0.628\\ \hline
\textbf{accumulo}&J48&SQ&0.155&0.407&\textbf{isis}&J48&SQ&0.011&0.606\\ \hline
\textbf{accumulo}&J48&TLCC&0.225&0.543&\textbf{isis}&J48&TLCC&0.024&0.399\\ \hline
\textbf{accumulo}&Logistic&CKJM&0.147&0.37&\textbf{isis}&Logistic&CKJM&0.011&0.489\\ \hline
\textbf{accumulo}&Logistic&R2C&0.104&0.509&\textbf{isis}&Logistic&R2C&0.008&0.602\\ \hline
\textbf{accumulo}&Logistic&R2RS&0.205&0.555&\textbf{isis}&Logistic&R2RS&0.009&0.723\\ \hline
\textbf{accumulo}&Logistic&SQ&0.173&0.364&\textbf{isis}&Logistic&SQ&0.014&0.535\\ \hline
\textbf{accumulo}&Logistic&TLCC&0.093&0.229&\textbf{isis}&Logistic&TLCC&0.045&0.023\\ \hline
\textbf{accumulo}&NaiveBayes&CKJM&0.134&0.147&\textbf{isis}&NaiveBayes&CKJM&0.019&0.264\\ \hline
\textbf{accumulo}&NaiveBayes&R2C&0.15&0.353&\textbf{isis}&NaiveBayes&R2C&0.009&0.223\\ \hline
\textbf{accumulo}&NaiveBayes&R2RS&0.208&0.609&\textbf{isis}&NaiveBayes&R2RS&0.008&0.876\\ \hline
\textbf{accumulo}&NaiveBayes&SQ&0.14&0.212&\textbf{isis}&NaiveBayes&SQ&0.019&0.269\\ \hline
\textbf{accumulo}&NaiveBayes&TLCC&0.129&0.089&\textbf{isis}&NaiveBayes&TLCC&0.059&0.034\\ \hline
\textbf{accumulo}&RandomForest&CKJM&0.135&0.554&\textbf{isis}&RandomForest&CKJM&0.01&0.554\\ \hline
\textbf{accumulo}&RandomForest&R2C&0.116&0.62&\textbf{isis}&RandomForest&R2C&0.009&0.57\\ \hline
\textbf{accumulo}&RandomForest&R2RS&0.244&0.593&\textbf{isis}&RandomForest&R2RS&0.009&0.622\\ \hline
\textbf{accumulo}&RandomForest&SQ&0.129&0.492&\textbf{isis}&RandomForest&SQ&0.01&0.558\\ \hline
\textbf{accumulo}&RandomForest&TLCC&0.205&0.538&\textbf{isis}&RandomForest&TLCC&0.021&0.414\\ \hline
\textbf{ignite}&Bagging&CKJM&0.801&0.593&\textbf{tika}&Bagging&CKJM&0.034&0.796\\ \hline
\textbf{ignite}&Bagging&R2C&0.358&0.128&\textbf{tika}&Bagging&R2C&0.025&0.604\\ \hline
\textbf{ignite}&Bagging&R2RS&0.773&0.904&\textbf{tika}&Bagging&R2RS&0.027&0.283\\ \hline
\textbf{ignite}&Bagging&SQ&0.802&0.537&\textbf{tika}&Bagging&SQ&0.03&0.738\\ \hline
\textbf{ignite}&Bagging&TLCC&0.646&0.499&\textbf{tika}&Bagging&TLCC&0.034&0.425\\ \hline
\textbf{ignite}&BayesNet&CKJM&0.768&0.554&\textbf{tika}&BayesNet&CKJM&0.041&0.875\\ \hline
\textbf{ignite}&BayesNet&R2C&0.308&0.175&\textbf{tika}&BayesNet&R2C&0.021&0.254\\ \hline
\textbf{ignite}&BayesNet&R2RS&0.79&0.56&\textbf{tika}&BayesNet&R2RS&0.001&0.008\\ \hline
\textbf{ignite}&BayesNet&SQ&0.761&0.535&\textbf{tika}&BayesNet&SQ&0.002&0.1\\ \hline
\textbf{ignite}&BayesNet&TLCC&0.695&0.424&\textbf{tika}&BayesNet&TLCC&0&0\\ \hline
\textbf{ignite}&J48&CKJM&0.77&0.648&\textbf{tika}&J48&CKJM&0.038&0.871\\ \hline
\textbf{ignite}&J48&R2C&0.402&0.156&\textbf{tika}&J48&R2C&0.068&0.575\\ \hline
\textbf{ignite}&J48&R2RS&0.758&0.941&\textbf{tika}&J48&R2RS&0.02&0.238\\ \hline
\textbf{ignite}&J48&SQ&0.78&0.514&\textbf{tika}&J48&SQ&0.017&0.563\\ \hline
\textbf{ignite}&J48&TLCC&0.643&0.516&\textbf{tika}&J48&TLCC&0.006&0.117\\ \hline
\textbf{ignite}&Logistic&CKJM&0.781&0.468&\textbf{tika}&Logistic&CKJM&0.032&0.758\\ \hline
\textbf{ignite}&Logistic&R2C&0.361&0.111&\textbf{tika}&Logistic&R2C&0.027&0.679\\ \hline
\textbf{ignite}&Logistic&R2RS&0.85&0.97&\textbf{tika}&Logistic&R2RS&0.022&0.317\\ \hline
\textbf{ignite}&Logistic&SQ&0.803&0.374&\textbf{tika}&Logistic&SQ&0.026&0.471\\ \hline
\textbf{ignite}&Logistic&TLCC&0.706&0.42&\textbf{tika}&Logistic&TLCC&0&0\\ \hline
\textbf{ignite}&NaiveBayes&CKJM&0.831&0.191&\textbf{tika}&NaiveBayes&CKJM&0.038&0.513\\ \hline
\textbf{ignite}&NaiveBayes&R2C&0.443&0.149&\textbf{tika}&NaiveBayes&R2C&0.082&0.721\\ \hline
\textbf{ignite}&NaiveBayes&R2RS&0.932&0.479&\textbf{tika}&NaiveBayes&R2RS&0.01&0.271\\ \hline
\textbf{ignite}&NaiveBayes&SQ&0.842&0.169&\textbf{tika}&NaiveBayes&SQ&0.026&0.596\\ \hline
\textbf{ignite}&NaiveBayes&TLCC&0.864&0.66&\textbf{tika}&NaiveBayes&TLCC&0.031&0.179\\ \hline
\textbf{ignite}&RandomForest&CKJM&0.801&0.614&\textbf{tika}&RandomForest&CKJM&0.031&0.754\\ \hline
\textbf{ignite}&RandomForest&R2C&0.35&0.119&\textbf{tika}&RandomForest&R2C&0.023&0.571\\ \hline
\textbf{ignite}&RandomForest&R2RS&0.761&0.93&\textbf{tika}&RandomForest&R2RS&0.021&0.229\\ \hline
\textbf{ignite}&RandomForest&SQ&0.798&0.565&\textbf{tika}&RandomForest&SQ&0.029&0.633\\ \hline
\textbf{ignite}&RandomForest&TLCC&0.642&0.521&\textbf{tika}&RandomForest&TLCC&0.024&0.387\\ \hline

\end{tabular}
\caption{Precision and recall when using single metrics families in predicting impacted classes}
\label{tab:Rq3P&R}
\end{table*}

\cleardoublepage

\begin{table*}[!t]
\label{tab:pre_rec_rq4}
\begin{tabular}{|l|l|l|l|l||l|l|l|l|l|}
\hline
\textbf{Dataset}  & \textbf{Classifier} & \textbf{MetricFamily} & \textbf{Precision} & \textbf{Recall} & \textbf{Dataset} & \textbf{Classifier} & \textbf{R2RS Usage} & \textbf{Precision} & \textbf{Recall} \\ \hline
\textbf{accumulo}&Bagging&Using&0.225&0.134&\textbf{isis}&Bagging&Using&0.197&0.117\\ \hline
\textbf{accumulo}&Bagging&Not Using&0.226&0.128&\textbf{isis}&Bagging&Not Using&0.104&0.148\\ \hline
\textbf{accumulo}&J48&Using&0.282&0.197&\textbf{isis}&J48&Using&0.108&0.173\\ \hline
\textbf{accumulo}&J48&Not Using&0.244&0.144&\textbf{isis}&J48&Not Using&0.11&0.204\\ \hline
\textbf{accumulo}&Logistic&Using&0.168&0.075&\textbf{isis}&Logistic&Using&0.076&0.013\\ \hline
\textbf{accumulo}&Logistic&Not Using&0.56&0.029&\textbf{isis}&Logistic&Not Using&0.041&0.013\\ \hline
\textbf{accumulo}&NaiveBayes&Using&0.376&0.398&\textbf{isis}&NaiveBayes&Using&0.026&0.228\\ \hline
\textbf{accumulo}&NaiveBayes&Not Using&0.231&0.076&\textbf{isis}&NaiveBayes&Not Using&0.026&0.111\\ \hline
\textbf{accumulo}&RandomForest&Using&0.28&0.165&\textbf{isis}&RandomForest&Using&0.18&0.084\\ \hline
\textbf{accumulo}&RandomForest&Not Using&0.263&0.153&\textbf{isis}&RandomForest&Not Using&0.196&0.098\\ \hline
\textbf{ignite}&Bagging&Using&0.864&0.951&\textbf{tika}&Bagging&Using&0.317&0.15\\ \hline
\textbf{ignite}&Bagging&Not Using&0.787&0.611&\textbf{tika}&Bagging&Not Using&0.063&0.3\\ \hline
\textbf{ignite}&J48&Using&0.846&0.896&\textbf{tika}&J48&Using&0.144&0.164\\ \hline
\textbf{ignite}&J48&Not Using&0.781&0.587&\textbf{tika}&J48&Not Using&0.076&0.286\\ \hline
\textbf{ignite}&Logistic&Using&0.803&0.659&\textbf{tika}&Logistic&Using&0.215&0.343\\ \hline
\textbf{ignite}&Logistic&Not Using&0.716&0.333&\textbf{tika}&Logistic&Not Using&0&0\\ \hline
\textbf{ignite}&NaiveBayes&Using&0.919&0.538&\textbf{tika}&NaiveBayes&Using&0.131&0.571\\ \hline
\textbf{ignite}&NaiveBayes&Not Using&0.858&0.232&\textbf{tika}&NaiveBayes&Not Using&0.046&0.507\\ \hline
\textbf{ignite}&RandomForest&Using&0.861&0.889&\textbf{tika}&RandomForest&Using&0.356&0.236\\ \hline
\textbf{ignite}&RandomForest&Not Using&0.8&0.716&\textbf{tika}&RandomForest&Not Using&0.091&0.357\\ \hline

\end{tabular}
\caption{Precision and recall when using or not using the R2RS family of metrics in predicting impacted classes}
\label{tab:Rq4P&R}
\end{table*}

%% file: Biography.tex
\section{Biography}
\label{sec:Biography}

\begin{wrapfigure}{l}{0.15\textwidth}
  \begin{center}
    \includegraphics[width=0.15\textwidth]{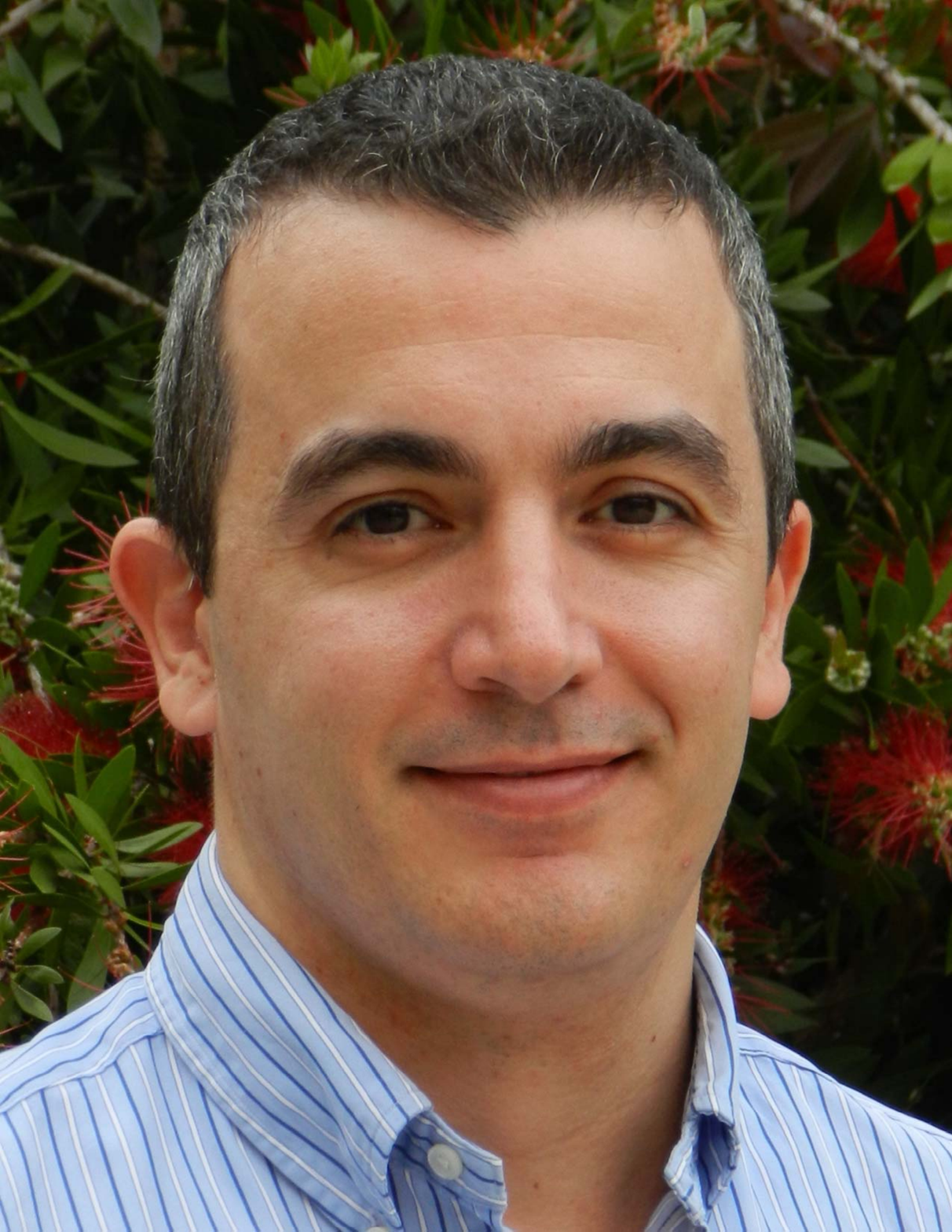}
  \end{center}
\end{wrapfigure} 
\textbf{Davide Falessi} is an Associate Professor of Computer Science at the California Polytechnic State University, USA. He has served in various positions for the community as: chair and technical program committee member for 50+ academic conferences, regular reviewer for 10+ academic journals, and reviewer for national research councils such as NSF, NSERC, and MIUR. He is the Associate Editor in Software Economics of IEEE Software and a senior member of IEEE. 
He is also a Member at Large of the IEEE Computer Society publications board. Falessi is a passionate  software engineering teacher. His main research interest is in devising and empirically assessing scalable solutions for the development of software intensive systems. He received the PhD, MSc, and BSc degrees in Computer Engineering from the University of
Rome Tor Vergata, Italy.
\vspace{1cm}

\begin{wrapfigure}{l}{0.15\textwidth}
  \begin{center}
    \includegraphics[width=0.15\textwidth]{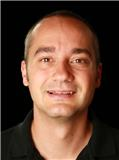}
  \end{center}
\end{wrapfigure} 
\textbf{Justin Roll} is a software development engineer at Amazon USA, working primarily on a distributed multi-tenant platform that helps power the amazon ecommerce website. He is currently interested in machine learning, distributed systems, and natural language processing. He received a BA from the University of California, Santa Cruz USA, and worked for Yardi Systems as a database administrator before obtaining his MSc degree in Computer Science from the California Polytechnic State University, USA
\vspace{1cm}

\begin{wrapfigure}{l}{0.15\textwidth}
  \begin{center}
    \includegraphics[width=0.15\textwidth]{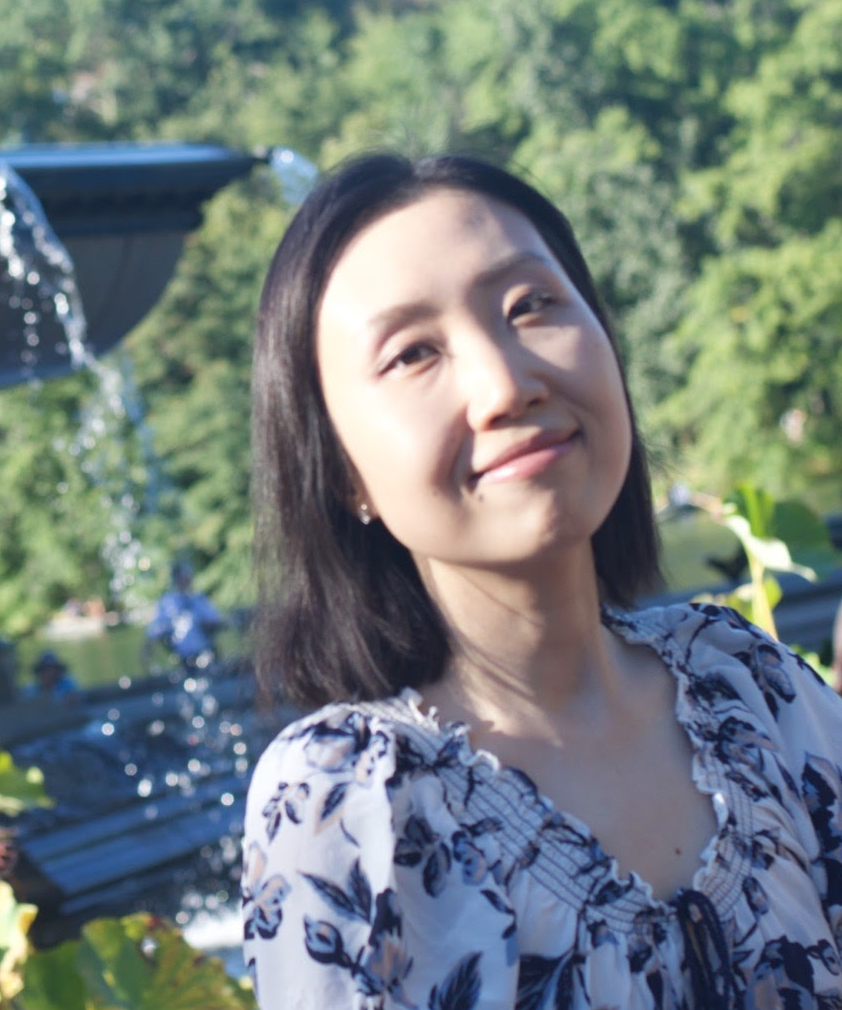}
  \end{center}
\end{wrapfigure} 
\textbf{Jin L.C. Guo} is an Assistant Professor in the School of Computer Science at McGill University. She received her Ph.D. degree in Computer Science and Engineering from the University of Notre Dame. Her research interests are software traceability, software maintenance and evolution, and human aspects of software engineering. Her current research focuses on utilizing Natural Language Processing techniques to construct connections within and across heterogeneous software engineering data.

\begin{IEEEbiography}[{\includegraphics[width=1in,height=1.25in,clip,keepaspectratio]{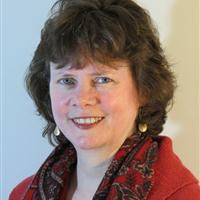}}]{Jane Cleland-Huang} is a Professor of Software Engineering at the University of Notre Dame. She received her PhD from the University of Illinois at Chicago. Her research interests focus upon Software and Systems Traceability for Safety- Critical Systems with a particular emphasis on the application of machine learning techniques to solve large-scale software and requirements engineering problems. She is Chair of IFIP 2.9 Working Group on Requirements Engineering, and has recently served as Associate Editor in Chief of Transactions on Software Engineering, and on the Editorial board of IEEE Software.
\end{IEEEbiography}